\documentclass[aps,preprint,floatfix]{revtex4} 

\usepackage[dvips]{epsfig}
\usepackage{amsfonts}
\usepackage{latexsym}

\newcommand{\tdiff}[2]{\frac{d #1}{d #2}}
\newcommand{\tdiffonly}[1]{\frac{d}{d #1}}
\newcommand{\pdiff}[2]{\frac{\partial #1}{\partial #2}}

\newcommand{\expval}[1]{\langle #1 \rangle}

\newcommand{\abs}[1]{\left| #1 \right|}

\newcommand{\bea}{\begin{eqnarray}}
\newcommand{\eea}{\end{eqnarray}}
\newcommand{\beann}{\begin{eqnarray*}}
\newcommand{\eeann}{\end{eqnarray*}}
\newcommand{\trace}[1]{{\rm Tr} \left\{ #1 \right\}}
\newcommand{\order}[1]{{\cal O}\left( #1 \right)}
\newcommand{\squeeze}[1]{\hat{\cal #1}}
\newcommand{\commute}[2]{\left[ #1 , #2 \right]}
\newcommand{\timo}{\hat{\cal T}}

\newcommand{\pro}{\widehat{\mathfrak P}}
\newcommand{\lio}{\widehat{\mathfrak L}}
\newcommand{\nn}{\nonumber}
 
\begin{document}  

\title{Dynamical Casimir Effect in a Leaky Cavity at Finite Temperature} 


\author{Gernot Schaller$^{1,*}$, 
Ralf Sch\"utzhold$^{1,2}$, 
G\"unter Plunien$^1$, 
and Gerhard Soff$^1$}


\affiliation{$^1$Institute for Theoretical Physics, 
Dresden University of Technology
D-01062 Dresden, Germany
\\
$^2$Department of Physics and Astronomy,
University of British Columbia,
Vancouver, B.~C., V6T 1Z1, Canada
\\
$^*$Electronic address : {\tt schaller@@theory.phy.tu-dresden.de}}
 
\date{\today} 

\begin{abstract}
  The phenomenon of particle creation within an almost resonantly
  vibrating cavity with losses is investigated for the example
  of a massless scalar field at finite temperature. A leaky cavity
  is designed via the insertion of a dispersive mirror into a larger ideal
  cavity (the reservoir).
  In the case of parametric resonance the rotating wave approximation
  allows for the construction of an effective Hamiltonian. The number 
  of produced particles is then calculated  
  using response theory as well as a non-perturbative approach. In
  addition we study the associated master equation and briefly discuss
  the effects of detuning. 
  The exponential growth of the particle numbers and the strong 
  enhancement at finite temperatures found earlier for ideal cavities
  turn out to be essentially preserved.
  The relevance of the results for experimental tests of quantum
  radiation via the
  dynamical Casimir effect is addressed. Furthermore the generalization 
  to the electromagnetic field is outlined.

\end{abstract}  

\maketitle




\section{Introduction}\label{Sintro} 

Since the pioneering work of Casimir \cite{casimir} the phenomena of
quantum field theory under the influence of external conditions have
attracted the interest of many authors, see e.g. \cite{bordag}.  
The original prediction by Casimir, i.e., the attractive force
generated between two perfectly conducting objects placed in the vacuum,
has been verified in different experimental setups with relatively high 
precision \cite{lamoreauxmohideenbressi}.
However, its dynamic counterpart with non-stationary boundary conditions
inducing interesting effects like the creation of particles out of the
vacuum has not yet been observed rigorously in a corresponding
experiment. The observation of quantum radiation could provide a 
substantial test of the foundations of quantum field theory and thus 
be of special relevance. Generally we understand the term quantum
radiation to denote the conversion of virtual quantum fluctuations
into real particles due to external disturbances. For the special case
of the external disturbances being moving mirrors this phenomenon is known
as the Dynamical Casimir effect.  

These striking effects have been investigated by several authors, for
an overview see e.g. \cite{bordag,jaekelreynaud} and references therein.
We will focus on the effect of particle creation within a constructed
-- resonantly vibrating -- leaky cavity. This case is of special importance
for an experimental verification of the dynamical Casimir effect since
the generation of particles is enhanced drastically by resonance
effects. 
Employing different methods and approaches it has already been shown
for ideal cavities (see e.g. \cite{finitetemp}) that under resonance 
conditions (i.e., when one of the boundaries 
performs harmonic oscillations at twice the frequency of one
of the eigenmodes of the cavity) the phenomenon of parametric
resonance (see e.g. \cite{jijungparksoh}) will occur. 
In the case of an ideal cavity (i.e., one with
perfectly reflecting mirrors) this is known to lead to an exponential 
growth of the resonance mode particle occupation numbers,
cf. \cite{finitetemp,dalvit,dodonov1,dodonov2,dodonov3}.
 
In view of this prediction an experimental observation of quantum
radiation using the dynamical Casimir effect appears to be rather 
simple -- provided the cavity is vibrating at resonance for a
sufficiently long period of time. However, this point of view is too naive
since neither ideal cavities do exist nor is it possible to
match the external frequency to the fundamental eigenfrequency of the
cavity with arbitrary precision.
Consequently, it is essential to include effects of leaks as well as
effects of detuning, see also \cite{golestanian}.
 
Investigations concerning effects of losses have been 
performed for example in \cite{jrrad} in 1+1
space-time dimensions based on conformal mapping methods as developed in
\cite{davies}. However, these considerations are a priori restricted to
1+1 dimensions and can not be obviously generalized to higher dimensions.
In 3+1 dimensions the character of the mechanism generating quantum 
radiation -- e.g., the resonance conditions -- 
differs drastically from the 1+1 dimensional situation.

More realistic (3+1 dimensional) cavities were considered in 
\cite{leakydodonov} where the
effects of losses were taken into account by virtue of a master
equation ansatz. However, this master equation had not been derived
starting from first principles. It has already been noted in
\cite{leakydodonov} that the employed ansatz is 
adequate for a stationary cavity -- but not necessarily for a
dynamic one. 
In addition, most papers did not include temperature effects -- which
may contribute significantly in an experiment. 
It has been shown in Ref. \cite{finitetemp} that for an ideal cavity
the effect of particle production at finite temperature is enhanced by several
orders of magnitude in comparison with the pure vacuum contribution. 

In this article we will adopt the canonical approach which has proven 
to be general, successful, and is -- in addition -- 
also capable of including temperature effects. 
However, the aforementioned approach still 
lacks a generalization for systems with losses.
We are aiming at providing a remedy in this field \cite{schaller}. 

This paper is organized as follows:
In Sec.~\ref{Sbasic} we present a model system and derive the
effective Hamiltonian for the resonance case. 
In Sec.~\ref{Squadresp} we will calculate the number 
of created particles in the cavity after one of the walls has performed 
resonant oscillations by means of response theory.
In Sec.~\ref{Smaster} we will derive and solve the associated master 
equation and show consistency 
with the results obtained in Sec.~ \ref{Squadresp}.
In Sec.~\ref{Snonperturb} a non-perturbative approach is presented
and compared with the other results. 
We derive a treshold condition - valid for leaky cavities - for a 
possible detuning from the fundamental resonance in Sec.~\ref{Sdetune} .
We shall close with a summary, a discussion, a conclusion and an outlook.

Throughout this paper natural units given by $\hbar=c=k_B=1$ will be used.


\section{General Formalism}\label{Sbasic}

\subsection{The Leaky Cavity}\label{SSsetup}

\begin{figure}
  \includegraphics[width=8cm]{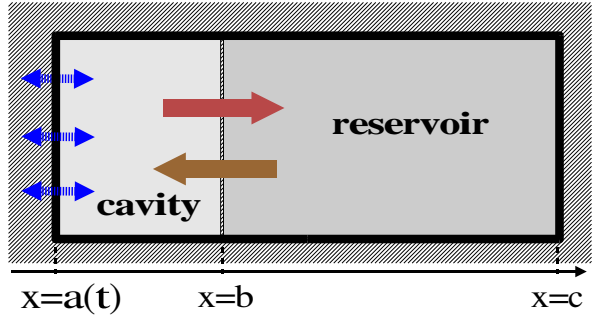}
  \caption{\label{Fsystem} Model of a leaky cavity. A large
  ideal cavity is split up by a dispersive mirror into a leaky cavity
  and a reservoir. The left (ideal) wall of the cavity is vibrating.}
\end{figure}

We want to investigate the effects of a non-ideal cavity in view
of the dynamical Casimir effect. For that purpose we have to
construct a suitable model system. One simple way to do that is to
insert a dispersive mirror into an ideal cavity while keeping
all other walls perfectly reflecting. Thereby two leaky cavities are
formed. Particles in the left imperfect
cavity are now able to leave into the right larger box (the reservoir).
For reasons of simplicity we consider a rectangular cavity as 
depicted in Fig.~\ref{Fsystem}.
The setup in Fig. \ref{Fsystem} is not a new idea. A similar -- but
static -- system has already been treated in \cite{scully,banachloche}.
However, here in addition the left wall is moving with a prescribed 
trajectory during the time interval $[0,T]$. For ideal cavities this is
known to lead to a squeezing of the vacuum state which causes the creation of
particles inside the cavity, see e.g. \cite{finitetemp}. 

Note that we are assuming a finite reservoir with a discrete spectrum 
instead of an infinite one leading to a continuum of modes. 
Since, in an experimental setup, the vibrating cavity will most likely be 
surrounded by walls, etc. (imposing additional boundary conditions), this 
assumption should be justified.

Assuming a surrounding perfectly reflecting wall is a first
idealization of the real situation. However, in order to minimize the error
obtained by this procedure the experiment could be designed in this way, see
also Fig. \ref{Fproposal} in section \ref{Sconclude} below.

\begin{figure}
  \includegraphics[width=8cm]{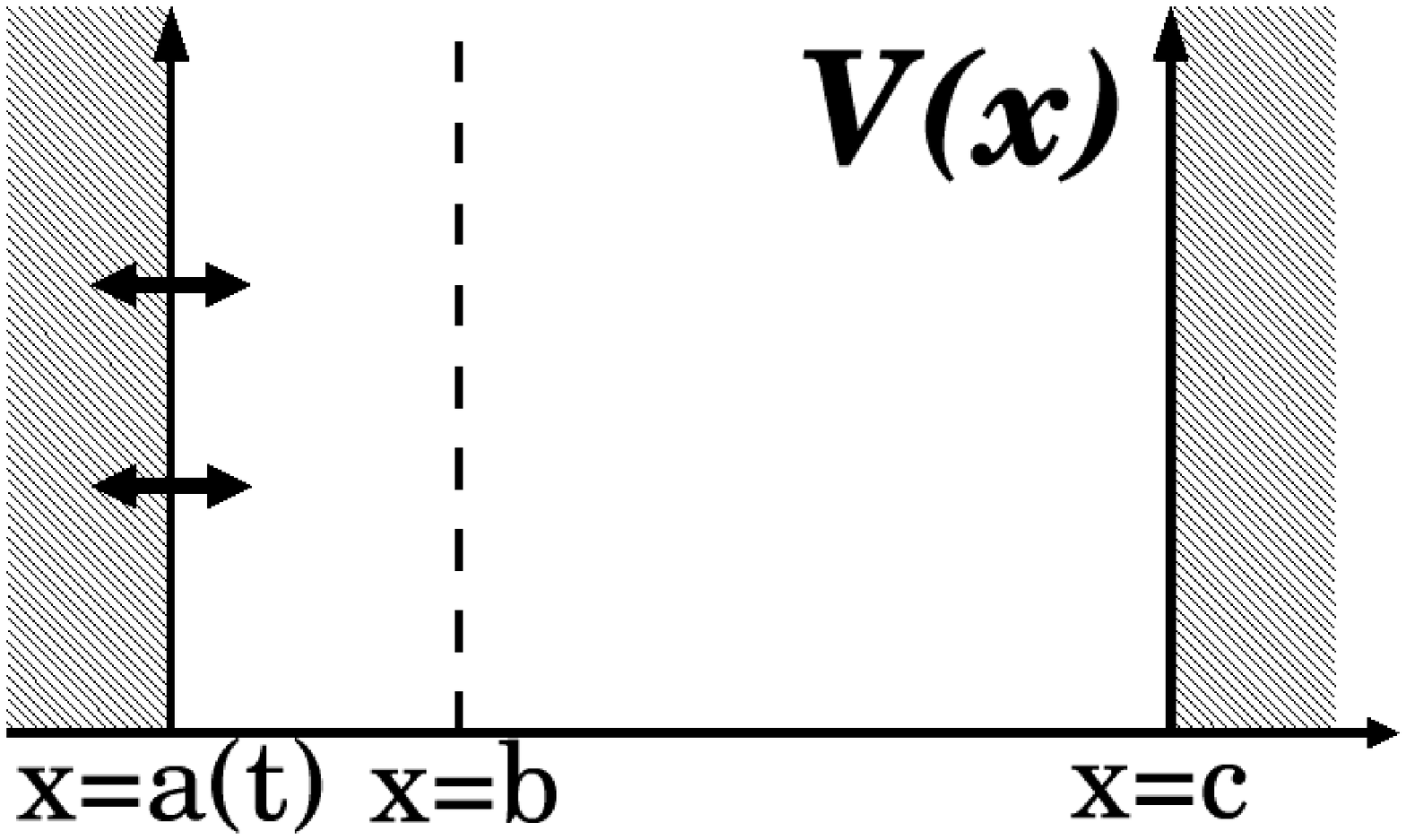}
  \caption{\label{Fpotential}
  Diagram of the $x$-dependence of the used potential.} 
\end{figure}

The ideal mirrors can be simulated by infinitely high potential walls
inducing Dirichlet boundary conditions.
For the additional dispersive mirror we use the $\delta$-type model potential
proposed in Ref. \cite{calogeracos,salomone}
\bea 
  V(x;t) = \left\{ 
         \begin{array}{ll} \gamma \delta(x-b)&\mbox{if $a(t)<x<c$}\\ 
                           \infty & \mbox{otherwise} 
        \end{array} 
      \right. \, ,
\eea 
see also Fig.~\ref{Fpotential}.
The parameter $\gamma$ represents the
transmittance of the internal mirror, whose reflection and
transmission amplitudes are determined as \cite{calogeracos}
\bea
  {\mathcal R} = -\frac{i \gamma}{\omega+i \gamma}\,, \qquad
  {\mathcal T} = \frac{\omega}{\omega+i \gamma}\,. 
\eea
Note that the general procedure presented in this article is
independent of the particular form of the potential -- the 
aforementioned one has just been chosen for convenience. For a more
realistic scenario one could apply square-well or Gaussian potentials.
In a realistic experiment where one would want to create photons 
instead of scalar particles a dispersive mirror could be realized 
using a thin dielectric slab with a very high dielectric constant. 
Such a mirror could then be approximated by a space-dependent
permittivity $\varepsilon (x) = 1 + \alpha \delta (x)$. 
This will be addressed in Sec.~ \ref{SSemfield}.

\subsection{Hamiltonian}\label{SSfield}

Throughout this article we will use the notation of \cite{tremblingcav}
where the particle production in an ideal vibrating cavity was 
calculated -- for a more general treatment see e.g. \cite{blackhole}. 
We consider a massless and neutral scalar field coupled to an external potential:
\bea\label{lagrangian}
  {\cal L} = \frac{1}{2} (\partial_\mu \Phi)(\partial^\mu \Phi) - V\Phi^2\,. 
\eea
The perfect mirrors can be incorporated by imposing the
corresponding boundary condition on $\Phi$.
By expanding the field $\Phi$
\bea
  \Phi (\mbox{\boldmath $r$},t) = \sum_\mu Q_\mu (t) 
        f_\mu (\mbox{\boldmath $r$};t)
\eea
into a complete and
orthonormal set of functions $f_\mu (\mbox{\boldmath $r$};t)$ 
satisfying
\bea\label{Efeatures}
  \int d^3 r f_\mu^*(\mbox{\boldmath $r$},t) 
  f_\nu(\mbox{\boldmath $r$},t) = \delta_{\mu\nu}\,,\\
  \sum_\mu f_\mu^*(\mbox{\boldmath $r$};t) 
  f_\mu(\mbox{\boldmath $r'$};t) = \delta^3
  (\mbox{\boldmath $r$}-\mbox{\boldmath $r'$})\,,\\
  \{2 V -\Delta\}f_\mu(\mbox{\boldmath $r$};t) = \Omega_\mu^2(t) 
  f_\mu(\mbox{\boldmath $r$};t)
\eea
one can reach a more convenient form suitable for
doing calculations. Since $\Phi$ is a real field, we can choose the
set $f_\mu$ to be real. Note that the time dependence of
eigenfunctions and eigenfrequency is solely induced by the moving boundary.
Inserting this expansion into Eq.~(\ref{lagrangian}) transforms 
the Lagrangian into \cite{tremblingcav} 
\bea
  L  = \int d^3 r\,{\cal L} & = & 
        \frac{1}{2} \sum_\mu \dot{Q}_\mu^2 
      - \frac{1}{2} \sum_\mu \Omega_\mu^2 (t) Q_\mu^2\nonumber\\
    &&+\sum_{\mu\nu} Q_\mu M_{\mu\nu}(t) \dot{Q}_\nu\nonumber\\
    &&+\frac{1}{2} \sum_{\mu\nu\kappa} Q_\mu M_{\mu\kappa}(t)
      M_{\nu\kappa}(t) Q_\nu\,,
\eea 
where $M_{\mu\nu}(t)$ is an antisymmetric matrix given by
\bea
   M_{\mu\nu} = \int d^3 r \pdiff{f_\mu}{t}(\mbox{\boldmath $r$};t)
        f_\nu (\mbox{\boldmath $r$};t).
\eea 
This matrix describes the coupling strength between two different modes. 
We introduce the canonical conjugated momenta
\bea 
  P_\mu = \pdiff{L}{\dot{Q}_\mu} 
        =\dot{Q}_\mu+\sum_\nu Q_\nu M_{\nu\mu}(t)\,.
\eea
Furthermore we apply the usual Legendre transform to a Hamiltonian 
representation and perform the quantization. This yields
\bea
  \hat{H} =\frac{1}{2} \sum_\mu \hat{P}_\mu^2
        +\frac{1}{2} \sum_\mu \Omega_\mu^2 \hat{Q}_\mu^2
        +\sum_{\mu\nu} \hat{P}_\mu M_{\mu\nu} \hat{Q}_\nu\,.
\eea
The above Hamiltonian can be sub-classified into
\bea
  \hat{H} = \hat{H}_0 + \hat{H}_I^S + \hat{H}_I^V\,,
\eea
where the single Hamiltonians are given by
\bea
  \hat{H}_0 & = & \frac{1}{2} \sum_\mu \hat{P}_\mu^2 
  + \frac{1}{2} \sum_\mu (\Omega_\mu^0)^2 \hat{Q}_\mu^2\,,\\
  \hat{H}_I^S & = & \frac{1}{2} \sum_\mu \Delta \Omega_\mu^2 (t) 
  \hat{Q}_\mu^2\,,\\ 
  \hat{H}_I^V & = & \sum_{\mu\nu} 
  \hat{P}_\mu M_{\mu\nu} (t) \hat{Q}_\nu\,.
\eea
The deviation $\Delta \Omega_\mu^2 (t) = \Omega_\mu^2
(t)-(\Omega_\mu^0)^2$ denotes the difference of the (squared) 
time-dependent eigenfrequencies $\Omega_\mu^2 (t)$ from the
unperturbed ones $(\Omega_\mu^0)^2$.
The first term $\hat{H}_0$ is the Hamiltonian of harmonic oscillators. The
remaining terms will further on be called squeezing interaction
Hamiltonian and velocity interaction Hamiltonian. We want to point out 
that in the case of a static system
(where the eigenfunctions $f_\mu$ and eigenfrequencies 
$\Omega_\mu$ are constant in time) the complete interaction Hamiltonian 
$\hat{H}_I= \hat{H}_I^S + \hat{H}_I^V$ will vanish. The derivation of 
the eigenfunctions $f_\mu (\mbox{\boldmath $r$},t)$ and 
$M_{\mu\nu}(t)$ will be treated in the following subsection.  


\subsection{Eigenmodes}\label{SSeigenmodes}

\begin{figure}
  \includegraphics[width=9cm]{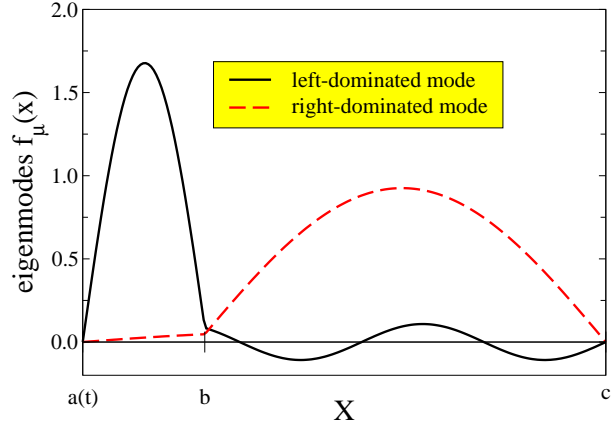}
  \caption{\label{Feigenmodes}
  Illustration of the lowest left- and right-dominated 
  eigenmodes $f_{1,r/l}^x(x)$ for $\eta_{1,l}=0.1$.} 
\end{figure}

As has already been mentioned, we want to find a set of functions
satisfying 
$\{2 V(\mbox{\boldmath $r$};t) -\Delta\}f_\mu(\mbox{\boldmath $r$};t) 
= \Omega_\mu^2(t) f_\mu(\mbox{\boldmath $r$};t)$.  
Any time dependence can only be induced by the moving boundaries.
At first we will just consider the spatial dependence i.e., the
stationary problem. 
The differential equation can be treated using the separation ansatz 
$f_\mu (\mbox{\boldmath $r$}\,) =
f_\mu (\mbox{\boldmath $r$}_{\|}) f_\mu (\mbox{\boldmath $r$}_{\bot})$
where $f_\mu (\mbox{\boldmath $r$}_{\|})$ depends only on
the coordinate parallel to the wall velocity and 
$f_\mu (\mbox{\boldmath $r$}_{\bot})$ is
dependent on the perpendicular coordinates. For the special case of
our model system this means
$f_\mu (\mbox{\boldmath $r$}\,) =
f_\mu^x (x) f_\mu^y (y) f_\mu^z (z)$ leading to the
trivial $y$ and $z$ dependence of the eigenfunctions
\bea
  f_\mu^y (y) & = & \sqrt{\frac{2}{\Delta y}} 
        \sin\left[\frac{n_y \pi}{\Delta y} y\right] \quad , \quad
        \Omega_\mu^y = \frac{n_y \pi}{\Delta y}\,,\\
  f_\mu^z (z) & = & \sqrt{\frac{2}{\Delta z}} 
        \sin\left[\frac{n_z \pi}{\Delta z} z\right] \quad , \quad
        \Omega_\mu^z = \frac{n_z \pi}{\Delta z}
\eea
with $\Delta y$ and $\Delta z$ denoting the dimensions of the cavity
and the frequencies relating via
\bea
  \Omega_\mu^2 = (\Omega_\mu^x)^2 + 
  (\Omega_\mu^y)^2+(\Omega_\mu^z)^2\,.
\eea
The remaining differential equation reads
\bea
  \left\{ 2 \gamma \delta (x-b) -\partial_x^2 \right\} f_\mu^x (x) 
  = (\Omega_\mu^x)^2 f_\mu^x (x)
\eea
where the Dirichlet boundary conditions coming from the perfect
mirrors on either side can be satisfied by the ansatz
\bea\label{Eansatz}
  f_\mu^x (x) = \left\{ 
         \begin{array}{ll} L_\mu \sin[\Omega_\mu^x (x-a)]  
         & \mbox{if $\;a<x<b$} \\ 
         R_\mu \sin[\Omega_\mu^x (c-x)] & \mbox{if $\;b<x<c$}\\
         0 & \mbox{elsewhere} 
        \end{array} 
        \right. \,. 
\eea
The eigenfunctions have to obey the continuity conditions
\cite{calogeracos} 
\bea
  f_\mu^x (x \downarrow b) - f_\mu^x (x \uparrow b) & = & 0\,,\\
  \pdiff{f_\mu^x}{x}(x\downarrow b)-\pdiff{f_\mu^x}{x}(x\uparrow b)
  & = & 2 \gamma f_\mu^x (b)\,,
\eea
where the latter can be obtained via integration.
These conditions can be combined to an eigenvalue equation for 
$\Omega_\mu^x$
\bea\label{Eeigenvalue}
  -\frac{2\gamma}{\Omega_\mu^x} = \cot\left[\Omega_\mu^x(b-a)\right] + 
   \cot\left[\Omega_\mu^x(c-b)\right] = -\frac{2}{\eta_\mu}\,. 
\eea
Though there is no obvious analytical solution of this equation, a
numerical solution can always be obtained for given cavity 
parameters $\{a,b,c,\gamma\}$. However, via
introducing the dimensionless perturbation parameter
$\eta^{}_\mu = \Omega_\mu^x/\gamma$ it is also possible to obtain 
an approximate analytical solution. 
Note that this parameter is small $\eta^{}_\mu \ll 1$ in the
limit of the internal mirror being nearly perfectly reflecting. 
Since the trigonometric functions are very sensitive to small
frequency variations one can solve the equation using a series
expansion in $\eta_\mu$. 
It is obvious that if the right hand side goes to $-\infty$ one of
the addends or even both can become relevant. This depends on the ratio
$(b-a)/(c-b)$ and its inverse which are both assumed to be non-integer 
numbers in the following non-perturbative calculations implying that 
only one of the addends is dominating.
Accordingly, expanding around the poles of one addend one yields a
polynomial that can be solved for $\Omega_\mu^x$ as a series expansion
in $\eta^{}_\mu \ll 1$.  
Depending on the chosen addend one obtains two sets of approximate 
eigenfrequencies
\bea\label{Eeigenfreq}
  \Omega_{n^{}_x, l}^x & = & \frac{n^{}_x \pi}{b-a} 
        -\frac{1}{2(b-a)}\eta^{}_{n^{}_x,l}\nn\\
        &&+\frac{1}{4(b-a)}\cot\left(n^{}_x\pi
        \frac{c-b}{b-a}\right)\eta^2_{n^{}_x,l}\nn\\
        &&+\order{\eta^3_{n^{}_x,l}}\,,\nn\\
  \Omega_{n^{}_x, r}^x & = & \frac{n^{}_x \pi}{c-b} 
        -\frac{1}{2(c-b)}\eta^{}_{n^{}_x,r}\nn\\
        &&+\frac{1}{4(c-b)}\cot\left(n^{}_x\pi
        \frac{b-a}{c-b}\right)\eta^2_{n^{}_x,r}\nn\\
        &&+\order{\eta^3_{n^{}_x,r}}\,,
\eea
which constitute a determining polynomial for $\Omega_\mu^x$. Note
that the index $\mu=(n^{}_x,l/r)$ is a multi-index, where $l$ and $r$
stand for left-dominated and right-dominated, respectively.
However, it can be shown easily that the quality of the linear (in $\eta$)
approximation suffices already for moderate values of $\gamma \ge 50$. 
The insertion of (\ref{Eeigenfreq}) into the ansatz (\ref{Eansatz}) 
leads to two classes of eigenfunctions: left-dominated and
right-dominated, respectively. The differences between those are 
clearly visible in Fig.~\ref{Feigenmodes}. 

In order to avoid the confusion arising from a set of perturbation 
parameters $\{\eta_\mu\}$ we will introduce the fundamental one via 
\bea
  \eta = \eta_{1l} = \frac{\Omega_{1l}^x}{\gamma} \,,
\eea
to which all others are evidently related via 
$\eta_\mu = \Omega_\mu^x/\Omega_{1l}^x \eta$. 
Note that this distinction between the classes of eigenfunctions 
is applicable only for small values of $\eta$.

Consequently, the eigenfunctions can be labeled by multi-indices 
$\mu = (n_x, n_y, n_z, r/l)$: 3 quantum numbers 
$n_{x,y,z} \in {\mathbb N}_{+}$ and a flag $r / l$
denoting the class (right- or left-dominated, respectively) of the 
eigenfunction.
Now we want to consider the effect of one moving boundary.
It is taken into account by substituting
$a \to a(t)$ everywhere in the eigenmodes and
-frequencies. Thereby a time dependence of the
eigenfunctions as well as of the eigenfrequencies is introduced.
This induces a non-vanishing coupling matrix 
$M_{\mu\nu}(t)$ as well as the frequency deviation
$\Delta\Omega_\mu^2(t)$. For small oscillations of the boundary
\bea
 a(t) = a_0 + \epsilon (b-a_0) \sin(\omega t)
\eea
with a small amplitude $\epsilon \ll 1$ it will be useful to 
separate the time dependence using
\bea
  M_{\mu\nu}(t) & = & \dot{a}(t) \int d^3 r 
  \pdiff{f_\mu}{a}(\mbox{\boldmath $r$};t)
  f_\nu (\mbox{\boldmath $r$};t)\nn\\
  & = & \dot{a}(t) m_{\mu\nu}(t) \,.
\eea
The geometry factor $m_{\mu\nu}(t)$ is approximately
constant $m_{\mu\nu}(t)=m_{\mu\nu}+\order{\epsilon}$ in this
case. Consequently, one is lead to
\bea
  M_{\mu\nu}(t) = m_{\mu\nu} \dot{a} (t) + \order{\epsilon^2}\,.
\eea
Since the time-dependence of the right-dominated modes is less 
complicated than that of the left-dominated ones, it is advantageous 
to exploit the antisymmetry of $M_{\mu\nu}$ which also implies an 
antisymmetry of $m_{\mu\nu}$. For the following calculations the 
coupling of the lowest left-dominated mode $\mu = (1, 1, 1, l)$ to some
right-dominated one $\nu=(n_x,n_y,n_z, r)$ will be of special 
relevance.
The $y$ and $z$ integrations simply generate Kronecker
symbols and therefore the geometry factor results as
\bea\label{Egeomfactor}
  m_{\mu,\nu} & = & - \delta_{1,n_y}\delta_{1,n_z} 
        \int_a^c dx f_{\mu}^x \pdiff{f_{\nu}^x}{a}\nn\\
        & = &
        \frac{\delta_{1,n_y}\delta_{1,n_z} n_x (-1)^{n_x}
        \sqrt{\mbox{$\displaystyle\frac{b-a}{c-b}$}}\,
        \mbox{$\displaystyle\frac{\Omega_{n_x r}^x}{\Omega_{1l}^x}$}}
        {(c-b)\sin\left(n_x\pi\mbox{$\displaystyle\frac{b-a}{c-b}$}\right)
        \left[n_x^2\left(\mbox{$\displaystyle\frac{b-a}{c-b}$}\right)^2
        -1\right]} \eta\nn\\
        &&+\order{\eta^2} \nn\\
        &=& \order{\eta} \,. 
\eea


\subsection{Canonical Quantization}\label{SScanonical}

Aiming at the calculation of possible particle creation effects
(expectation values of particle number operators) it is
convenient to introduce the creation and annihilation operators
\bea
  \hat{a}_\mu (t)= \frac{1}{\sqrt{2\Omega_\mu^0}}
        \left(\Omega_\mu^0 \hat{Q}_\mu (t)+ i
        \hat{P}_\mu (t) \right)\, ,
\eea
obeying the usual bosonic equal time commutation relations 
\bea \label{Ecommrel}
  \commute{\hat{a}_\mu(t)}{\hat{a}_\nu^\dagger(t)} &=& 
  \delta_{\mu\nu}\,,\nn\\
  \commute{\hat{a}_\mu(t)}{\hat{a}_\nu(t)\phantom{^\dagger}} &=& 0\,,\nn\\
  \commute{\hat{a}_\mu^\dagger(t)}{\hat{a}_\nu^\dagger(t)} &=& 0 \,.
\eea
These operators diagonalize the free Hamiltonian
\bea 
  \hat{H}_0 = \sum_\mu \Omega_\mu^0 \left( 
  \hat{a}_\mu^\dagger(t)
  \hat{a}_\mu(t) + \frac{1}{2} \right)\,.
\eea
The following calculations will most conveniently be done in the 
interaction picture where the dynamics of an observable is governed
by $\hat{H}_0$
\bea
  \tdiff{\hat{Y}}{t} &  = & i
  \commute{\hat{H}_0}{\hat{Y}} 
  + \left(\pdiff{\hat{Y}}{t}\right)_{\rm explicit}\,.
\eea
For reasons of generality and to include finite temperature effects we
describe the state of a quantum system by a statistical operator whose
dynamics is determined by the von Neumann equation
\bea
  \tdiff{\hat{\rho}}{t} &  = & -i
  \commute{\hat{H}_I}{\hat{\rho}}\,.
\eea
Note that this equation without any explicit time dependence
$(\partial\hat{\rho}/\partial{t})_{\rm explicit}$ 
leads to an unitary time evolution, see also 
\cite{finitetemp} and Sec.~\ref{Smaster}.

In this picture the time dependence of the creation and annihilation 
operators turns out to be
\bea\label{Eladderoscillation}
  \hat{a}_\mu (t)  = \hat{a}_\mu e^{-i\Omega_\mu^0 t} \,.
\eea
However, this trivial time dependence gives rise to the possibility of
parametric resonance which enhances the chances to verify the effect
of particle creation experimentally.
Further-on we will denote the initial creation and annihilation
operators by $\hat{a}_\mu (0) = \hat{a}_\mu$.
Note that in this picture the particle number operator 
$\hat{N}_\mu=\hat{a}_\mu^\dagger \hat{a}_\mu$ 
is time independent for all modes.


\subsection{Rotating Wave Approximation}\label{SSrwa}

In the interaction picture the time-evolution operator is given by
\bea\label{Etime}
  \hat{U}(T,0) = \timo_t \exp\left[-i\int_{0}^{T} \left(
                \hat{H}_I^S (t)+\hat{H}_I^V (t) \right) dt\right]
\eea
where $\timo_t$ denotes time-ordering. If the interaction Hamiltonian
$\hat{H}_I$ leads to small corrections, the usual procedure is to apply 
perturbation theory via expanding the exponential.
Evidently, this would imply that $\hat{U}$ is close to the identity. 
On the other hand, in order to make an experimental verification of 
quantum radiation feasible, the time-evolution operator should deviate 
significantly from the identity.
Therefore a different approximation needs to be found.
For the case of parametric resonance this may be accomplished via
applying the rotating wave approximation (RWA), see
e.g. \cite{finitetemp,law,wuchuleung}. Within this scenario the left 
boundary performs harmonic oscillations obeying 
$a(t) = a_0 + \epsilon(b-a_0) \sin(\omega t)$ with the dimensionless 
amplitude\footnote{For uniqueness we restrict ourselves to $\epsilon>0$.} 
$\epsilon \ll 1$ and the external vibration frequency 
$\omega$  during the time interval $[0,T]$.
This also implies an oscillating time-dependence of the frequency 
deviation and coupling matrix
\bea
  \Delta\Omega_\mu^2 (t) & = & 2\Omega_\mu^0 \pdiff{\Omega_\mu^0}{a_0}
        (b-a_0) \epsilon \sin(\omega t) +\order{\epsilon^2}\,,\\
  M_{\mu\nu}(t) & = & m_{\mu\nu} (b-a_0) \omega \epsilon \cos(\omega t)
        +\order{\epsilon^2}\,,
\eea 
see also Sec.~\ref{SSeigenmodes}.
Together with the trivial time-dependence of the ladder operators 
(\ref{Eladderoscillation}) in the interaction picture this enables us to 
perform the RWA. Expanding the time evolution operator (\ref{Etime})
into an infinite series one can treat the time-ordering terms in the
following way: In analogy to
\bea
  \timo_t \left[ \hat{H}_I(t_1) \hat{H}_I(t_2) \right]
  & = & \Theta(t_2-t_1)\commute{\hat{H}_I(t_2)}{\hat{H}_I(t_1)} \nn\\
  && + \hat{H}_I(t_1) \hat{H}_I(t_2)
\eea
one can rewrite all these terms to yield a multiple product of 
Hamiltonians without time-ordering and terms involving commutators 
with Heaviside step functions. In the resonance case, i.e., when
$\omega = 2 \Omega_\mu^0$ the terms with commutators can yield nothing 
but strongly oscillating integrands \cite{finitetemp}, as can also be 
seen by Fourier-expanding the involved Hamiltonians. 
With the duration of the perturbation being
sufficiently long, i.e., with $\omega T \gg 1$, the contribution of
these terms to (\ref{Etime}) will be comparably small. 
As a consequence, time-ordering can be neglected to all
orders within the RWA.

The remaining integrals -- without time-ordering -- factorize and can 
be resummated to yield an effective time evolution operator
\bea
  \hat{U}_{\rm eff}(T,0) \stackrel{\rm RWA}{=} 
        \exp\left(-i\hat{H}_{\rm eff}^I T\right)\,,
\eea
where the effective interaction Hamiltonian
\bea
  \hat{H}_{\rm eff}^I T \stackrel{\rm RWA}{=} \int_0^T \left[\hat{H}_I^S (t) 
        +\hat{H}_I^V (t)\right] dt
\eea
still has to be calculated.
The above time integration involves many oscillating terms. 
Again, with the duration of the vibration being
sufficiently long $\omega T \gg 1$, i.e., after many oscillations, 
the time-integrated interaction Hamiltonian can be approximated in 
the following way:
Since the time average of purely oscillating
terms is rather small compared to that of constant contributions we
may neglect the former ones. As a result, in the series expansion of
the time-evolution operator $\hat{U}(T,0)$ only those terms where the 
oscillation of the ladder operators is compensated by the external
vibrations -- represented by $\Delta \Omega_\mu^2 (t)$ and 
$M_{\mu\nu}(t)$, respectively -- will be kept. 
Strictly speaking, in the above equation terms of 
$\order{\epsilon^I (\omega T)^J}$ are neglected by the RWA 
if $I>J$ holds. The terms with $J=K$ -- i.e., exactly the
terms in which the oscillations of the creation and annihilation
operators (\ref{Eladderoscillation}) are compensated by the external 
time dependence $\Delta\Omega_\mu^2(t)$ and $M_{\mu\nu}(t)$] -- will be
kept. (Note that terms with $J<K$ do not occur.) 

The general squeezing interaction Hamiltonian reads
\bea
  \hat{H}_I^S & = &\sum_\mu \frac{1}{2}
        \pdiff{\Omega_\mu^0}{a_0} \epsilon (b-a_0) \sin(\omega t)
        \left[ (\hat{a}_\mu)^2 (t) +(\hat{a}_\mu^\dagger)^2 (t)\right.\nn\\
        &&\left.+\hat{a}_\mu (t) \hat{a}_\mu^\dagger (t)
        +\hat{a}_\mu^\dagger (t) \hat{a}_\mu (t)\right]
        +\order{\epsilon^2}\,.
\eea
Accordingly, within the RWA only the terms fulfilling the squeezing
resonance condition, see also e.g.
\cite{leakydodonov,finitetemp,quantumrad,braginsky} 
\bea
  \omega=2\Omega_\mu^0
\eea
will be kept. In general $\mu$ can also be a right-dominated mode, but
note that in this case the effective squeezing Hamiltonian would be of
$\order{\eta^2}$, since according to (\ref{Eeigenfreq}) the
right-dominated eigenfrequencies do not depend on $a(t)$ up to
$\order{\eta}$. Therefore we will restrict ourselves to left-dominated
modes $\mu$ and among those in particular to the lowest one, i.e., as
commonly done we consider the case of fundamental resonance
\bea\label{Erescond1}
  \omega=2\Omega_{111l}^0=2\Omega_L^0\,.
\eea
From now on this mode will be abbreviated by the index $L=(1,1,1,l)$
throughout this publication. 
Consequently, by virtue of
\bea
  \frac{1}{T} \int_0^T \sin(\omega t) e^{\pm i \omega t} dt
        \stackrel{\rm RWA}{=} \pm \frac{i}{2}
\eea
an effective squeezing Hamiltonian can be derived
\bea\label{Eeffhs}
  \hat{H}_{\rm eff}^S = 
  i \xi \left[(\hat{a}_L^\dagger)^2-(\hat{a}_L)^2\right]\,,
\eea
where $\xi$ is given by 
\bea
  \xi = \frac{1}{4}\epsilon\Omega_{L}^0
  \left(\frac{\Omega_{L}^{x 0}}{\Omega_{L}^0}\right)^2 \,.
\eea 
Obviously $\hat{H}_{\rm eff}^S$ is a generator for a
squeezing operator for the mode $L$ with $\xi$ being the squeezing
parameter.

The same procedure can be applied for the velocity interaction
Hamiltonian which reads in the vibration case 
\bea
  \hat{H}_I^V & = &\frac{i}{2}\sum_{\mu\nu}
  \sqrt{\frac{\Omega_\mu^0}{\Omega_\nu^0}}m_{\mu\nu} 
  \epsilon\omega (b-a_0) \cos(\omega t)\times\nn\\  
  &&\left[\hat{a}_\mu^\dagger (t)\hat{a}_\nu^\dagger (t) 
  +\hat{a}_\mu^\dagger (t)\hat{a}_\nu (t)\right.\nn\\
  &&\left.-\hat{a}_\mu (t)\hat{a}_\nu^\dagger (t)
  -\hat{a}_\mu (t)\hat{a}_\nu (t)\right]+\order{\epsilon^2}\,.
\eea
However, the occurrence of inter-mode couplings now results in a different
resonance condition (see also \cite{finitetemp,quantumrad})
\bea\label{Eresvelocity}
  \omega & = & \abs{\Omega_{\mu}^0 \pm \Omega_{\nu}^0} \,. 
\eea
Depending on the frequency spectrum of the cavity under consideration this
resonance condition might be fulfilled by several pairs $\mu\nu$, but
here we will assume for simplicity that only one such pair exists. 
Though in any case via
\bea
  \frac{1}{T}\int_0^T \cos(\omega t) e^{\pm i\omega t} dt
        \stackrel{\rm RWA}{=} \frac{1}{2}
\eea
an effective velocity Hamiltonian can be derived, two major
distinctions should be made. 

\paragraph{\underline{$\oplus$ coupling}} $\omega=\Omega_1^0 + \Omega_2^0$.
  In this case one yields an effective velocity Hamiltonian given by
  \bea
  \hat{H}_{\rm eff}^V & = & \frac{i}{8}
        \left(\sqrt{\frac{\Omega_1^0}{\Omega_2^0}}
        -\sqrt{\frac{\Omega_2^0}{\Omega_1^0}}\right)
        m_{12}\epsilon\omega(b-a_0)\times\nn\\  
        &&\left(\hat{a}_1^\dagger \hat{a}_2^\dagger 
        -\hat{a}_1\hat{a}_2\right)\,,
  \eea
  which is a non-diagonal multi-mode squeezing Hamiltonian.
  Note that if one wants to fulfill squeezing and velocity resonance 
  conditions simultaneously 
  ($2 \Omega_L^0 = \omega = \Omega_1^0 + \Omega_2^0$), the
  number of possible combinations reduces significantly, since then the
  velocity resonance condition can not be fulfilled by two distinct 
  left-dominated modes. For reasons of brevity we do not consider this
  case here.

\paragraph{\underline{$\ominus$ coupling}} $\omega=\Omega_2^0 - \Omega_1^0$.
  Here the resulting effective velocity Hamiltonian does not
  resemble a squeezing but a hopping operator
  \bea
    \hat{H}_{\rm eff}^V & = & \frac{i}{8}
        \left(\sqrt{\frac{\Omega_1^0}{\Omega_2^0}}
        +\sqrt{\frac{\Omega_2^0}{\Omega_1^0}}\right)
        m_{12}\epsilon\omega(b-a_0)\times\nn\\  
        &&\left(\hat{a}_1^\dagger \hat{a}_2 
        - \hat{a}_1\hat{a}_2^\dagger\right)\,.
  \eea
  This coupling is of special interest since if one does not insist
  on simultaneously fulfilling both resonance conditions --
  parametric resonance might perhaps still be induced by lower external
  frequencies $\omega=\Omega_2^0-\Omega_1^0<2\Omega_L^0$ whose
  generation would be simpler in an experiment, see also
  Sec.~\ref{SSremarks}. 
  In the case of simultaneously fulfilling both conditions several 
  combinations may arise.
  \begin{itemize}
    \item The frequencies $\Omega_i^0$ both belong to either right- or
    left-dominated modes. In \cite{dalvit} it has been shown that for an
    ideal cavity with special dimensions, e.g. a cubic one, a strong 
    inter-mode coupling can occur. In that case a much smaller
    exponential particle creation rate has been found. Therefore in view of 
    an experimental verification this case is counterproductive and
    not considered here. Instead we propose a cavity with
    transcendental ratios of the dimensions such that there is no
    resonant inter-mode coupling of similar-dominated modes. In any
    case such a coupling would certainly require large quantum numbers
    of the involved modes.
    \item The frequency $\Omega_2^0$ represents a right-dominated mode 
    and $\Omega_1^0$ some left-dominated mode, respectively. 
    The lowest possible
    right-dominated frequency $\Omega_2=\Omega_R^0$ 
    would then be obtained when $\Omega_1^0=\Omega_L^0$. 
    As an example, this case will be
    considered here. We want to stress that the used
    methods are nevertheless applicable to any possible combination of
    couplings. 
  \end{itemize}   
Note that the situation would be completely different in 
1+1 space-time dimensions where -- due to the equidistant spectrum -- 
the velocity term always contributes, see e.g. 
\cite{finitetemp,dodonov1,dodonov2,dodonov3,dodonov4,dodonov5,dodonov6}. 
The coupling right-dominated mode fulfilling 
$\Omega_{n_x, n_y, n_z,r}=3\Omega_L^0$ will further-on be denoted with
the index $R=(n_x, n_y, n_z, r)$. Accordingly, in our considerations the
velocity Hamiltonian reads
\bea\label{Eeffhv}
  \hat{H}_{\rm eff}^V & = & i \chi  
        \left(\hat{a}_{L}^\dagger \hat{a}_{R}
        -\hat{a}_{L}\hat{a}_{R}^\dagger\right)
\eea
with
\bea
  \chi & = & \frac{1}{4} \epsilon \Omega_{L}^0
        \left(\sqrt{\frac{\Omega_{R}^0}{\Omega_L^0}}
        +\sqrt{\frac{\Omega_{L}^0}{\Omega_R^0}}\right)
        (b-a_0) m_{L,R} 
\eea
being the velocity parameter of the system. Since 
$\chi=\order{\epsilon \Omega_L^0 m_{L,R}}=\order{\epsilon\Omega_L^0\eta}$ 
[see also (\ref{Egeomfactor})]
it follows that $\chi/\xi=\order{\eta}\ll 1$ in the limiting case
of a nearly perfectly reflecting 
mirror\footnote{Note that for inter-mode coupling between modes of the
same region -- the corresponding scenario has been assumed in
\cite{dalvit} for an ideal cavity where one only has one region -- 
the velocity parameter $\chi$ would be of $\order{1}$ instead of 
$\order{\eta}$.}. This hopping operator is consistent with the visual 
picture of a semitransparent mirror. 


\subsection{Response Theory}\label{SSresponse}

We assume our system to be initially in a state of thermal equilibrium 
that can be described by the canonical ensemble
\bea
  \hat{\rho}(t=0)=\hat{\rho}_0 = \frac{\exp\left(-\beta \hat{H}_0\right)}
                {\trace{\exp\left(-\beta \hat{H}_0\right)}}
\eea
with $\beta$ denoting the initial inverse temperature.
The expectation value of an explicitly time-independent operator 
$\hat{Y}$ at time $t=T$ is given by
\bea
  \expval{Y(T)} & = & \trace{\hat{Y} \hat{\rho}(T)} \nn\\ 
                & = & {\rm Tr} 
        \left\{ \hat{Y} \timo_{t} 
        \exp\left(-i\int_0^T \hat{H}_I (t_1) dt_1\right)\right.\times\nn\\ 
                &   & \left. \hat{\rho}_0 \timo_{t}^\dagger  
                      \exp\left(+i\int_0^T \hat{H}_I (t_2)
  dt_2\right)\right\} 
\eea
where $\timo_t^\dagger$ denotes the anti-chronological operator
(anti-time ordering)  -- for a more
involved discussion see e.g. \cite{finitetemp}.

If the interaction Hamiltonian represented a small correction, one 
could expand the time-evolution operator into a perturbation series 
yielding a series expansion for $\expval{P(T)}$. However, for the 
resonance case this procedure is not justified:
In the rotating wave approximation in Sec.~\ref{SSrwa} the interaction
Hamiltonian simplified to
\bea\label{Eheff}
  \hat{H}_{\rm eff}  =  
                i \xi \left[(\hat{a}_L^\dagger)^2-(\hat{a}_L)^2\right] 
                + i \chi \left[ \hat{a}_L^\dagger
                \hat{a}_R-\hat{a}_L\hat{a}_R^\dagger \right] 
\eea
which implies for the time-evolution operator
\bea
  \hat{U}(T,0) \stackrel{\rm RWA}{=} 
  \exp\left(-i \hat{H}_{\rm eff}^I T\right)\,.
\eea
The whole expression for computing the expectation value of an operator
now becomes much simpler
\bea\label{Eexpval0}
  \expval{Y(T)} & \stackrel{\rm RWA}{=} &  
                {\rm Tr} \left\{\hat{Y} 
                \exp\left(-i(\hat{H}_{\rm eff}^S+\hat{H}_{\rm 
                eff}^V)T \right) \hat{\rho}_0\right.\times\nn\\ 
                &   & \left. 
                \exp\left(+i(\hat{H}_{\rm eff}^S+\hat{H}_{\rm
  eff}^V)T\right)\right\}\,, 
\eea
but since the correction $\hat{H}_{\rm eff}^S T$ is not small in the
case of interest above expression is still not practical for applying 
perturbation theory. Exploiting the smallness of the velocity
Hamiltonian it will prove useful to separate the two Hamiltonians. 
This can be achieved with the ansatz
\bea
  \exp\left(-i(\hat{H}_{\rm eff}^S+\hat{H}_{\rm eff}^V)\tau\right) 
  =  \exp\left(-i \hat{H}_{\rm eff}^S \tau\right) \hat{\sigma}(\tau)
\eea
with $\hat{\sigma}$ being an auxiliary operator. Differentiation with respect
to $\tau$ yields a differential equation that can be solved for
$\hat{\sigma}$ using the initial condition $\hat{\sigma}(0)={\bf 1}$. 
Introducing the parameter ordering $\timo_\tau$ in analogy to time 
ordering ($\timo_t$) the solution for $\hat{\sigma}$ can be cast 
into the form
\bea
  \hat{\sigma}(T) = \timo_{\tau} \left[\exp\left( - i \int_0^T 
                      \squeeze{H}_{\rm eff}^V(\tau) d\tau \right)\right]\,.
\eea
Here the squeezed effective velocity Hamiltonian has been introduced
\bea
  \squeeze{H}_{\rm eff}^V(\tau) = \exp\left(+i \hat{H}_{\rm eff}^S \tau\right) 
  \hat{H}_{\rm eff}^V \exp\left(-i \hat{H}_{\rm eff}^S \tau\right)  
\eea
which is now dependent on the parameter $\tau$.
Further-on we shall denote squeezed operators by calligraphy letters.
By inserting above equations into the expectation value
(\ref{Eexpval0}) one yields 
\bea
  \expval{Y(T)} & \stackrel{\rm RWA}{=} &  
                      {\rm Tr} \left\{ \squeeze{Y} (T) \timo_{\tau}  
                      \exp\left( - i \int_0^T 
                      \squeeze{H}_{\rm eff}^V(\tau_1) d\tau_1\right)
                        \right.\times\nn\\ 
                  &   & \left. \rho_0 \timo_{\tau}^\dagger  
                      \exp\left(+i\int_0^T \squeeze{H}_{\rm eff}^V 
                      (\tau_2) d\tau_2\right)\right\}.
\eea 
Please note that in this representation also the observables are
squeezed
\bea
  \squeeze{Y} (T) = \exp\left(+i \hat{H}_{\rm eff}^S T\right) 
  \hat{Y} \exp\left(-i \hat{H}_{\rm eff}^S T\right)
\eea
but here using the physical perturbation time $T$.
We will refer to this picture as
the squeezing interaction picture. Unfortunately the parameter ordering is
reintroduced by this procedure but as the advantage of these
manipulations we are now able to expand the 
expectation value $\expval{Y(T)}$ into a perturbation series with powers of 
$\squeeze{H}_{\rm eff}^V$. Keeping only terms to second order one finds
\bea\label{Eexpval}
  \expval{Y(T)} & = &\trace{\squeeze{Y} (T) \hat{\rho}_0}\nn\\ 
        & + & \trace{\squeeze{Y} (T) \commute{\hat{\rho}_0}{i \int d\tau_1  
               \squeeze{H}_{\rm eff}^V (\tau_1)}}\nn\\ 
        & + & \trace{\squeeze{Y} (T) \int d\tau_1 
               \squeeze{H}_{\rm eff}^V (\tau_1)\hat{\rho}_0 
               \int d\tau_2 \squeeze{H}_{\rm eff}^V (\tau_2)}\nn\\ 
        & - & \frac{1}{2}\trace{\squeeze{Y} (T) \timo_\tau \int d\tau_1  
               \squeeze{H}_{\rm eff}^V (\tau_1) \int d\tau_2  
               \squeeze{H}_{\rm eff}^V (\tau_2) \hat{\rho}_0}\nn\\ 
        & - & \frac{1}{2}\trace{\squeeze{Y} (T) \hat{\rho}_0 
               \timo_\tau^\dagger 
               \int d\tau_1 \squeeze{H}_{\rm eff}^V (\tau_1)  
               \int d\tau_2 \squeeze{H}_{\rm eff}^V (\tau_2)} \nn\\
        & + & \order{(\squeeze{H}_{\rm eff}^V)^3}
\eea
which is now a practical expression for calculating expectation values.


\section{The Quadratic Response}\label{Squadresp}

\subsection{Squeezing}\label{SSsqueeze}

According to the results of Sec.~\ref{SSresponse} in the squeezing
interaction picture 
both particle number operator and the effective velocity 
Hamiltonian have to be squeezed. The squeezing operator
\bea
  \hat{S}(\tau) = \exp\left(+i \hat{H}_{\rm eff}^S \tau \right) 
  = \exp\left(\xi
  \left[ (\hat{a}_L)^2-(\hat{a}_L^\dagger)^2\right]\tau\right)
\eea
implies the following well-known transformation rules (see
e.g. \cite{mandelwolf}) 
\bea
  \hat{b}_L (\tau) & = & 
                \hat{S} (\tau) \hat{a}_L \hat{S}^\dagger (\tau)\nn\\
                & = & \hat{a}_L \cosh (2\xi \tau) + \hat{a}_L^\dagger \sinh
                (2\xi \tau)\,,\\
  \hat{b}_L^\dagger (\tau) & = & 
                \hat{S} (\tau) \hat{a}_L^\dagger
                \hat{S}^\dagger (\tau)\nn\\ 
                & = & \hat{a}_L^\dagger \cosh (2\xi \tau) + \hat{a}_L \sinh
                (2\xi \tau)\,,
\eea
which can also be envisaged as a Bogoliubov transformation of the
ladder operators. Due to the commutation relations (\ref{Ecommrel}) 
other modes than the fundamental resonance mode $L$ are not affected 
by squeezing. Inserting the above expressions into the effective 
velocity Hamiltonian (\ref{Eeffhv})
one can easily derive its squeezed counterpart
\bea\label{EHeffvsqueezed}
  \squeeze{H}_{\rm eff}^V (\tau)
        & = & \hat{S} (\tau) \hat{H}_{\rm eff}^V \hat{S}^\dagger (\tau)\nn\\
        & = & i\chi \cosh(2\xi \tau)
        \left[\hat{a}_L^\dagger \hat{a}_R - \hat{a}_L
        \hat{a}_R^\dagger \right]\nn\\
        &   & + i\chi \sinh(2\xi \tau)
        \left[\hat{a}_L \hat{a}_R - \hat{a}_L^\dagger
        \hat{a}_R^\dagger \right]\,.
\eea
Note that the squeezed effective velocity Hamiltonian is now dependent
on the parameter $\tau$ and that it is still of
$\order{\chi}=\order{\eta}$ which justifies a perturbation-like treatment.

The same can be done for the particle number operators where again only the
fundamental resonance mode 
$\hat{N}_L=\hat{a}_L^\dagger \hat{a}_L$ is affected
\bea\label{ENLsqueezed}
  \squeeze{N}_{L} (T) & = & \hat{S} (T) \hat{N}_{L} \hat{S}^\dagger (T)
                        =\hat{S} (T) \hat{a}_L^\dagger \hat{S}^\dagger (T)
                        \hat{S} (T) \hat{a}_L \hat{S}^\dagger (T)\nn\\
                & = & \left[1+2\sinh^2(2 \xi T)\right]
                \hat{a}_L^\dagger \hat{a}_L\nn\\
                & & + \frac{1}{2}\sinh(4 \xi T)
                \left[(\hat{a}_L^\dagger)^2+(\hat{a}_L)^2\right]\nn\\
                & & + \sinh^2(2\xi T)\,.
\eea
For brevity we will denote the hyperbolic functions by 
\bea
  {\cal C}(T) & = & \cosh(2\xi T)\,,\nn\\
  {\cal S}(T) & = & \sinh(2\xi T)
\eea
throughout this paper.
                

\subsection{Expectation Values}\label{SSexpval}

Since we are mainly interested in the phenomenon of quantum radiation 
and thus therefore in the calculation of the cavity particle content after a time
$T$ when the disturbance has ended. Due to the dynamical disturbance the 
system leaves the thermodynamic equilibrium, see also
\cite{finitetemp}. 
The quadratic response of the expectation value of the particle  
number operator can be
calculated via substituting $\hat{Y}\to\hat{N}$ in equation
(\ref{Eexpval}). 
However, since the initial statistical operator $\hat{\rho}_0$ involves
arbitrarily high powers in $\hat{H}_0$, it is practical to rewrite 
the expression obtained from (\ref{Eexpval}). Utilizing the 
invariance of the trace under cyclic permutation and the property 
of time ordering
\bea
  \left\{ \hat{A}(t), \hat{B}(t') \right\}_{+} & = & 
  \hat{A}(t) \hat{B}(t')+\hat{B}(t') \hat{A}(t)\nn\\
  & = & \timo \left[ \hat{A}(t) \hat{B}(t') \right]\nn\\ 
  & & + \timo^\dagger \left[ \hat{A}(t) \hat{B}(t') \right]
\eea
the expectation value of interest can be cast into the more convenient form 
\bea\label{Eexpval1}
  \expval{N(T)} & = & \trace{\squeeze{N} (T) \hat{\rho}_0}\nn\\
        & + &  \trace{\commute{i \int d\tau_1 
                \squeeze{H}_{\rm eff}^V(\tau_1)}{\squeeze{N} (T)} 
                \hat{\rho}_0}\nn\\ 
        & + & {\rm Tr}\left\{\hat{\rho}_0 \int d\tau_2 
                \squeeze{H}_{\rm eff}^V (\tau_2) 
                \commute{\squeeze{N} (T)}{\int d\tau_1 
                \squeeze{H}_{\rm eff}^V
                (\tau_1)}\right.\nn\\ 
        & + & \left.\frac{1}{2} \hat{\rho}_0 \commute{\timo_\tau 
              \int d\tau_1 \squeeze{H}_{\rm eff}^V (\tau_1) 
              \int d\tau_2 \squeeze{H}_{\rm eff}^V (\tau_2)}
              {\squeeze{N} (T)}\right\}\nn\\
        & + & \order{(\squeeze{H}_{\rm eff}^V)^3}\,. 
\eea
This form is now suitable for evaluating the traces since all
commutators only concern a finite number of creation or annihilation
operators. Thus the quadratic response can be brought in
relation with the initial particle contents of the cavity
(Bose-Einstein distributions)
\bea 
  N_L^0 & = & \trace{\hat{a}_L^\dagger \hat{a}_L \hat{\rho}_0}= 
  \frac{1}{e^{\beta \Omega_L^0} -1}\,,\\
  N_R^0 & = & \trace{\hat{a}_R^\dagger \hat{a}_R \hat{\rho}_0}=
  \frac{1}{e^{\beta \Omega_R^0} -1}\,,
\eea
where $\beta$ stands for the initial inverse temperature of the system. These
mean occupation numbers incorporate the whole temperature
dependence of the quantum radiation -- as long as the back-reaction of
the field on the moving mirror can be neglected.
As we shall show later, the lowest order term ($\eta=0$) is 
in agreement with the results of an ideal cavity, as was 
considered for example in \cite{finitetemp,dalvit}. 
Also, since $\squeeze{H}_{\rm eff}^V$ contains only odd powers of creation
and annihilation operators for a single mode, the linear response
vanishes. Generally, every trace involving an odd power of ladder operators
vanishes and $\squeeze{N}$ as well as $\hat{\rho}_0$ do only contain even
powers.
Hence the last trace in Eq.~(\ref{Eexpval1}) constitutes the quadratic
answer. In contrast to an ideal cavity the terms with time ordering
are here especially important since they will be found to produce
leading order terms.


\subsection{Particle Creation}\label{SScreation1}

Using the squeezed operators (\ref{EHeffvsqueezed}) and
(\ref{ENLsqueezed}) it is now straightforward to compute the 
commutators and the traces in the expectation value (\ref{Eexpval1}). 
As a result one finds for particles in the fundamental resonance mode $L$
\bea\label{Equadleftresp}
  \expval{N_{L}(T)}     & = & {\cal S}^2(T)\nn\\ 
        &   & + \left[1+ 2 {\cal S}^2(T) \right]N_{L}^0\nn\\ 
        &   & + \frac{\chi^2}{4\xi^2} \left[3 {\cal C}^2(T) - 2 {\cal C}(T) -1 
                -2 \xi T {\cal S}(2T)\right]\nn\\ 
        &   & + \frac{\chi^2}{4\xi^2} \left[4 {\cal C}^2(T) - 2{\cal C}(T)-2
                -4 \xi T {\cal S}(2T)\right] N_{L}^0\nn\\ 
        &   & + \frac{\chi^2}{4\xi^2} \left[2{\cal C}^2(T)-2{\cal C}(T)\right]
        N_{R}^0\nn\\
        &   & + \order{\eta^3}\,.  
\eea
As was anticipated, the lowest order term 
${\cal S}^2(T) + [1+ 2 {\cal S}^2(T) ]N_{L}^0$ is in agreement with the 
results obtained in \cite{finitetemp} for an ideal cavity. The linear 
response (in $\eta$) vanishes. It might be of interest that the
leading terms $T{\cal S}(2T)$ in the quadratic answer stem from the 
time-ordering which is therefore very important.
One can see that at long
disturbance times $T$ these leading terms show
the failure of the quadratic approximation since the particle number
would become negative at some point. This is due to the fact that
(\ref{Eexpval1}) is a perturbation series in $\int_0^T \hat{H}_{\rm
eff}^V (t) dt$ which will always become large at some time $T$.
This problem can only be solved by including all orders in $\eta$, see
also Sec.~\ref{Snonperturb}.

Of course (\ref{Eexpval1}) can also be applied to the corresponding coupling
right-dominated mode (whose particle number operator is invariant
under squeezing) where one finds
\bea\label{Equadrightresp}
  \expval{N_{R}(T)} & = & N_R^0\nn\\ 
  &   & + \frac{\chi^2}{4\xi^2} 
  \left[2 {\cal C}^2(T) - 2 {\cal C}(T) +1\right]\nn\\ 
  &   & + \frac{\chi^2}{4\xi^2} 
  \left[2 {\cal C}^2(T) - 2{\cal C}(T)\right] N_{L}^0\nn\\ 
  &   & + \frac{\chi^2}{4\xi^2} 
  \left[-2{\cal C}(T)+2\right] N_{R}^0\nn\\
  &   & + \order{\eta^3}  \,.
\eea
Again the linear answer is vanishing. For $\eta=0$ there would 
not be any created particles in the reservoir due
to the dynamical Casimir effect corresponding to a
perfect internal mirror.

It is remarkable that the coefficient of $N_L^0$ in
$\expval{N_{R}(T)}$ equals the coefficient of $N_R^0$ 
in $\expval{N_{L}(T)}$. As we shall see in Sec.\ref{SScreation3},
this feature is preserved to all orders in $\eta$.


\section{The Master Equation Approach}\label{Smaster}

In this section it is our aim to derive the associated master equation 
for an effective statistical operator accounting for
the left leaky sub-cavity or left-dominated modes, respectively. 
So far (3+1) dimensional vibrating leaky cavities have only been treated
in different setups -- see e.g. \cite{leakydalvit} --
where the vibrating mirror is understood as a (quantized) harmonic oscillator 
coupled to the cavity field (the reservoir) or with master equations adequate 
rather for  stationary systems -- see e.g. \cite{leakydodonov}. 
It was assumed in \cite{leakydodonov} that these master equations
could also be applied when one of the boundaries was moving. 
The possibility of limitations to that procedure as well as the need
for a {\em rigorous derivation of the master equation for 
resonantly excited systems} have already been expressed in 
\cite{leakydodonov}.
We want to derive such an equation starting from first principles. 
As a test we will also solve 
the obtained master equation and recalculate the quadratic answer 
for the left mode particles to compare with the previous results of
Sec.~\ref{SScreation1}. To obtain a master equation we will 
closely follow the derivation given in \cite{mandelwolf}.


\subsection{Derivation of a Master equation}\label{SSmasterderive}

Throughout this section we will deploy the squeezing interaction
picture where not only the time dependence induced by $\hat{H}_0$ 
but also the dependence resulting from $\hat{H}_{\rm eff}^S$ is
determining the operator time evolution has already been proposed in
Sec.~\ref{SSresponse}. In this picture the time evolution of the statistical
operator is governed by a modified von Neumann equation
\bea\label{liouville}
  \pdiff{\hat{\rho}(t)}{t}      = -i \commute{\squeeze{H}_{\rm
                                eff}^V(t)}{\hat{\rho}(t)}
                                = -i \lio (t) \hat{\rho} (t)\,.
\eea
The above equation defines the action of the Liouvillian super 
operator $\lio$ on $\hat{\rho}$ (see also Ref. \cite{fick}). 
By defining the projection super operator $\pro$ via
\bea
  \pro \hat{Y} = \hat{\rho}_R (0) {\rm Tr}_R \left\{ \hat{Y} \right\}
\eea
for all observables $\hat{Y}$, where ${\rm Tr}_R$ means taking the 
trace solely over the right dominated modes we can introduce a reduced 
density operator accounting for the left-dominated modes only
\bea
  \hat{\rho}_L (t) & = & {\rm Tr}_R \left\{ \hat{\rho} (t) \right\}\,.
\eea
Combining above equations it can be shown \cite{mandelwolf}
that the dynamics of the full statistical operator $\hat{\rho}$ 
is governed by the Zwanzig master equation
\bea
  \pro \pdiff{\hat{\rho}(t)}{t} & = & - \pro \lio(t) \int_0^t \widehat{\mathfrak U}
  (t,t')(1-\pro)\lio(t')\pro\hat{\rho}(t')dt'\nn\\
  &&-i \pro \lio(t) \widehat{\mathfrak U} (t,0) (1-\pro) \hat{\rho}(0)\nn\\
  &&-i \pro \lio(t) \pro \hat{\rho}(t)
\eea
where
\bea
  \widehat{\mathfrak U} (t,t') =  
  \exp \left(-i (1-\pro)\int_{t'}^t \lio(t'') dt''\right) 
\eea
is the reduced time-evolution super operator. 
The Zwanzig master equation is exact to all orders in $\eta$ but
usually too complicated to be solved.
However, assuming an initial thermal equilibrium state and taking into
account that initially our system and reservoir do not interact (no 
correlations) it can be simplified considerably:

\begin{itemize}

\item[1.] In analogy to the argumentation concerning the vanishing of
the linear response in Sec.~\ref{SSexpval} it follows that 
\bea
  {\rm Tr}_R\left\{\squeeze{H}_{\rm eff}^V \hat{\rho}_R (0)\right\} =0
\eea
since $\squeeze{H}_{\rm eff}^V$ contains only odd and $\hat{\rho}_R (0)$
only even powers of the creation and annihilation operators for
the mode $R$. This can equivalently be written as
\bea 
  \pro\lio(t)\pro\hat{\rho}(t)=0\,.
\eea

\item[2.] In our setup the initially stationary system (stationary walls) 
does not permit interactions between system and reservoir, since both 
$M_{\mu\nu}(t_0)$ and $\Delta\Omega^2(t_0)$ will vanish. Consequently, assuming
thermal equilibrium, system and reservoir initially constitute independent 
subsystems which cannot be correlated, i.e., the initial statistical operator 
of the cavity modes factorizes
\bea 
  \hat{\rho}_0 = \hat{\rho} (0) = 
  \hat{\rho}_L (0) \otimes \hat{\rho}_R (0)\, ,
\eea
hence one finds (with ${\rm Tr}_R\{\hat{\rho}_R\}=1$)
\bea
  (1-\pro) \hat{\rho} (0) = 0\,.
\eea

\end{itemize}
These assumptions yield a simplified Zwanzig master equation 
\bea\label{Ezwanzigmastersimple}
  \pdiff{\pro \hat{\rho}(t)}{t}  =  
   - \pro \lio(t) \int_0^t \widehat{\mathfrak U}
                (t,t')\lio(t')\pro\hat{\rho}(t')dt'
\eea
which is exact but still too complicated to be solved. 

In order to gain a solvable equation we will apply further 
approximations:

\begin{itemize}
\item[a.]{\underline{Born approximation}}

Since $\lio=\order{\eta}$ one can approximate the reduced time-evolution 
operator via $\widehat{\mathfrak U} (t,t') = {\bf 1} + \order{\eta}$.
This neglects terms of $\order{\eta^3}$ if inserted into
(\ref{Ezwanzigmastersimple}) and yields
\bea
  \pdiff{\pro \hat{\rho}(t)}{t}  =  - \pro \lio(t) \int_0^t 
                \lio(t')\pro\hat{\rho}(t')dt' +\order{\eta^3}\,.
\eea
By employing the reduced density operator  
$\hat{\rho}_L(t)={\rm Tr}_R\left\{\hat{\rho}(t)\right\}$
one can equivalently write
\bea\label{E001}
  \pdiff{\hat{\rho}_L (t)}{t} & = & - {\rm Tr}_R\left\{ \lio(t) \int_0^t 
                \lio(t')\hat{\rho}_R (0) \hat{\rho}_L(t')dt'\right\}\nn\\
                & & +\order{\eta^3}\,.
\eea
This equation governs the time evolution of the effective
statistical operator $\hat{\rho}_L$ accounting for the left cavity.

\item[b.]{\underline{Markov approximation}}

The retardation in equation (\ref{E001}), i.e., the occurrence of
$\hat{\rho}_L(t')$, complicates the calculation of
$\hat{\rho}_L(t)$. Iterative application of (\ref{E001}) implies that
$\hat{\rho}_L(t')=\hat{\rho}_L(t)+\order{\eta^2}$.
Accordingly, we apply the Markov approximation, which is
also known as {\em short memory approximation}, simply by replacing 
$\hat{\rho}_L(t')\to\hat{\rho}_L(t)$ on the 
right hand side. Since 
$\lio=\order{\eta}$ we thereby neglect terms of $\order{\eta^4}$ and 
yield the Born-Markov master equation
\bea
  \pdiff{\hat{\rho}_L(t)}{t} & = & - {\rm Tr}_R\left\{ \int_0^t \lio(t) 
                \lio(t')\hat{\rho}_R (0)
  \hat{\rho}_L(t)dt'\right\}\nn\\
  & &+\order{\eta^3}\,,
\eea
thus having maintained the level of accuracy.
\end{itemize}
Using the definition of the Liouville operator $\lio$ in
Eq.~(\ref{liouville}) one can equivalently write
\bea
  \pdiff{\hat{\rho}_L(t)}{t} & = & 
                + {\rm Tr}_R\left\{ \squeeze{H}_{\rm eff}^V (t)
                \hat{\rho}_R(0)\hat{\rho}_L(t) 
                \int_0^t \squeeze{H}_{\rm eff}^V (t')dt'\right\}\nn\\
                & & - {\rm Tr}_R\left\{\squeeze{H}_{\rm eff}^V (t)
                \int_0^t \squeeze{H}_{\rm eff}^V (t')dt'
                \hat{\rho}_R(0)\hat{\rho}_L(t)\right\}\nn\\
                & & + {\rm h.c.} +\order{\eta^3}\,.
\eea
Finally, having evaluated both traces and after having performed the $t'$
integrations with the aid of (\ref{EHeffvsqueezed}), one yields the 
following master equation
\bea\label{Emaster}
   \pdiff{\hat{\rho}_L (t)}{t} & = &  f_1 (t)  
        \left[ 2 \hat{a}_L^\dagger\hat{\rho}_L (t) \hat{a}_L 
        -\hat{a}_L\hat{a}_L^\dagger\hat{\rho}_L (t) 
        -\hat{\rho}_L (t) \hat{a}_L \hat{a}_L^\dagger \right]\nn\\ 
        & & + f_2 (t) 
        \left[ 2 \hat{a}_L\hat{\rho}_L (t) \hat{a}_L^\dagger 
        -\hat{a}_L^\dagger \hat{a}_L\hat{\rho}_L (t) 
        -\hat{\rho}_L (t) \hat{a}_L^\dagger \hat{a}_L\right]\nn\\ 
        & & + f_3 (t) 
        \left[ \hat{a}_L^\dagger \hat{\rho}_L (t) \hat{a}_L^\dagger 
        + \hat{a}_L \hat{\rho}_L (t) \hat{a}_L \right]\nn\\ 
        & & - f_4 (t) 
        \left[ (\hat{a}_L^\dagger)^2 \hat{\rho}_L (t) 
        + \hat{\rho}_L (t) (\hat{a}_L)^2\right]\nn\\ 
        & & - f_5 (t) 
        \left[ (\hat{a}_L)^2 \hat{\rho}_L (t) 
        + \hat{\rho}_L (t) (\hat{a}_L^\dagger)^2\right]\nn\\ 
        & &+ \order{\eta^3}  
\eea
where the functions $f_i(t)$ are given by
\bea
  f_1(t) & = & \frac{\chi^2}{2\xi} {\cal S}(t)
                \left\{ {\cal C}(t)(2N_R^0+1)-N_R^0-1\right\}\,,\nn\\
  f_2(t) & = & \frac{\chi^2}{2\xi} {\cal S}(t)
                \left\{ {\cal C}(t)(2N_R^0+1)-N_R^0\right\}\,,\nn\\
  f_3(t) & = & \frac{\chi^2}{2\xi} 
                \left\{ {\cal C}^2(t)+{\cal S}^2(t)-{\cal C}(t)\right\}
                (2N_R^0+1)\,,\nn\\
  f_4(t) & = & \frac{\chi^2}{2\xi} 
                \left\{ \left[ {\cal C}^2(t)+{\cal S}^2(t)
                -{\cal C}(t)\right] N_R^0+{\cal C}^2(t)
                -{\cal C}(t)\right\}\,,\nn\\
  f_5(t) & = & \frac{\chi^2}{2\xi} 
                \left\{ \left[ {\cal C}^2(t)+{\cal S}^2(t)
                -{\cal C}(t)\right] N_R^0 +{\cal S}^2 (t)\right\} \,.
\eea
Via averaging over the degrees of freedom of the reservoir and by
applying the Born-Markov approximation we have now rigorously
derived a differential equation for an effective statistical operator
$\hat{\rho}_L(t)$ accounting for the leaky cavity.
This effective statistical operator
obeys a non-unitary time-evolution (changing entropy).
There are several possibilities to check the obtained master equation:
As the simplest tests one can verify that the time evolution 
preserves\footnote{Note however, that under certain conditions 
(e.g. $N_R^0=0$) the positive definiteness of the density operator may not be 
preserved -- if one applies the master equation (\ref{Emaster}) for large times
beyond its region of validity.}
the hermiticity and the trace of $\hat{\rho}_L$.
 
A better indication for a correct master
equation is the fact that if one takes the limit of no squeezing,
i.e., in this coupling $\xi \to 0$, the resulting equation corresponds to a
harmonic oscillator coupled to a thermal bath: With
\bea
  \lim_{\xi\to 0} f_1 (t) & = & \chi^2 t N_R^0\,,\\
  \lim_{\xi\to 0} f_2 (t) & = & \chi^2 t (N_R^0+1)\,,\\
  \lim_{\xi\to 0} f_{i=3,4,5} (t) & = & 0
\eea
one arrives at a simplified equation
\bea\label{Eredmaster}
  \pdiff{\hat{\rho}_L}{t} & \stackrel{\xi \to 0}{=} &
        \gamma_D  \frac{N_R^0}{2}   
        \left[ 2 \hat{a}_L^\dagger\hat{\rho}_L \hat{a}_L
        -\hat{a}_L\hat{a}_L^\dagger\hat{\rho}_L  
        -\hat{\rho}_L \hat{a}_L \hat{a}_L^\dagger \right]\nn\\ 
        && + \gamma_D \frac{N_R^0+1}{2}
        \left[ 2 \hat{a}_L\hat{\rho}_L  \hat{a}_L^\dagger 
        -\hat{a}_L^\dagger \hat{a}_L\hat{\rho}_L  
        -\hat{\rho}_L \hat{a}_L^\dagger \hat{a}_L\right]\nn\\
        &&+\order{\eta^3}\,. 
\eea
Apart from the time dependence of the damping coefficient 
$\gamma_D=2 \chi^2 t$ above equation is exactly the well-known 
master equation for a harmonic oscillator
coupled to a thermal bath, see e.g. \cite{amohandbook}.
 
The time-dependence of $\gamma_D$ is a remnant of the dynamic master
equation describing the time-dependent system in the unphysical limit
$\xi\to 0$.
However, in order to have a stronger indication for the correctness 
of our ansatz we want to solve the master equation
(\ref{Emaster}) explicitly.


\subsection{Approximate solution of the master equation}\label{SSmastersolve}

So far we have neglected terms of $\order{\eta^3}$. 
The functions $f_i(t)$ are already of $\order{\eta^2}$ which makes it
possible to maintain the level of accuracy by applying the
additional approximation $\hat{\rho}_L (t) \approx \hat{\rho}_L (0)$
on the right hand side of equation (\ref{Emaster}), 
which could also be envisaged as an additional Markov approximation.  
Accordingly, one is now able to yield a solution for $\hat{\rho}_L$
\bea\label{Edensity}
   \hat{\rho}_L (T) & = & \hat{\rho}_L(0)\nn\\ 
        & + & F_1 (T)  
        \left[ 2 \hat{a}_L^\dagger\hat{\rho}_L (0) \hat{a}_L 
        -\hat{a}_L\hat{a}_L^\dagger\hat{\rho}_L (0) 
        -\hat{\rho}_L (0) \hat{a}_L \hat{a}_L^\dagger \right]\nn\\ 
        & + & F_2 (T) 
        \left[ 2 \hat{a}_L\hat{\rho}_L (0) \hat{a}_L^\dagger 
        -\hat{a}_L^\dagger \hat{a}_L\hat{\rho}_L (0) 
        -\hat{\rho}_L (0) \hat{a}_L^\dagger \hat{a}_L\right]\nn\\ 
        & + & F_3 (T) 
        \left[ \hat{a}_L^\dagger \hat{\rho}_L (0) \hat{a}_L^\dagger 
        + \hat{a}_L \hat{\rho}_L (0) \hat{a}_L \right]\nn\\ 
        & - & F_4 (T) 
        \left[ (\hat{a}_L^\dagger)^2 \hat{\rho}_L (0) 
        + \hat{\rho}_L (0) (\hat{a}_L)^2\right]\nn\\ 
        & - & F_5 (T) 
        \left[ (\hat{a}_L)^2 \hat{\rho}_L (0) 
        + \hat{\rho}_L (0) (\hat{a}_L^\dagger)^2\right]\nn\\ 
        & + & \order{\eta^3}
\eea
with $F_i(T)=\int_0^T f_i(t) dt$. 
Given this effective statistical
operator for the leaky cavity one is now able to calculate the
number of created particles in all left-dominated modes. 
Note that for considering right-dominated
modes one would have to derive a statistical operator for the
reservoir.


\subsection{Particle Creation}\label{SScreation2}

Since we were working in the squeezing interaction picture -- where the
observables have to be squeezed -- the expectation value of the 
particle number operator reads
\bea
  \expval{N_L (T)} = 
        {\rm Tr}_L \left\{ \squeeze{N}_L (T) \hat{\rho}_L(T) \right\}\,.
\eea
Other left-dominated modes than the fundamental resonance mode $L$
are trivial to solve: Due to the commutation relations
(\ref{Ecommrel}) their ladder operators commute with those 
of the resonance mode $L$. This implies (due to the invariance of the trace
under cyclic permutations) that all higher order traces must vanish
and one just yields the trivial result of their initial occupation numbers. 
Inserting the approximate reduced density operator obtained in
Eq.~(\ref{Edensity}) as well as $\squeeze{N}_L$ into the above equation,
one can see immediately that zeroth and first order in $\eta$ agree with the
previous results but showing this for the second order is a bit
tedious. After some algebra one finally finds
complete agreement which the previous result found in Eq.
(\ref{Equadleftresp}) of Sec.~\ref{SScreation1} thus giving a 
strong indication for the validity of our
master equation within the RWA approach.


\subsection{Comparison with other results}\label{SScomparision}

In \cite{leakydodonov} the effects of losses are taken into account by
a generalized version of the simple master equation ansatz
\bea\label{Edodonov}
  \tdiff{\hat{\rho}}{t} = 
        i\commute{\hat{\rho}}{\hat{H}}
        +\frac{\gamma_D}{2}\left[2\hat{a}\hat{\rho}\hat{a}^\dagger
        -\hat{a}^\dagger\hat{a}\hat{\rho}
        -\hat{\rho}\hat{a}^\dagger\hat{a}\right]\,.
\eea
However, as we have observed in the previous calculations, this
master equation does not adequately describe the leaky cavity under
consideration:
\begin{itemize}
\item
It is restricted to the case where the initial state of the reservoir
is just the vacuum state and therefore does not include temperature
effects. This has been taken into account in \cite{leakydodonov}.
\item
In addition, even the master equation for an harmonic oscillator in a
thermal bath \cite{amohandbook}
\bea
  \pdiff{\hat{\rho}}{t} & = & 
        i\commute{\hat{\rho}}{\hat{H}}
        + \frac{\overline{n}}{2} \gamma_D   
        \left[ 2 \hat{a}^\dagger\hat{\rho} \hat{a}
        -\hat{a}\hat{a}^\dagger\hat{\rho}  
        -\hat{\rho} \hat{a}  \hat{a}^\dagger \right]\nn\\ 
        && + \frac{\overline{n}+1}{2} \gamma_D  
        \left[ 2 \hat{a}\hat{\rho}  \hat{a}^\dagger 
        -\hat{a}^\dagger \hat{a}\hat{\rho}  
        -\hat{\rho} \hat{a}^\dagger \hat{a}\right] 
\eea
cannot be assumed to describe the system correctly. Even if one
identifies the Hamiltonian $\hat{H}$ in above equation with the
effective squeezing Hamiltonian $\hat{H_{\rm eff}^S}$ this master
equation goes along with serious problems since the Markov
approximation is not justified anymore. This complication reflects the
inherent dynamic character of our system.
As we have shown in
Sec.~\ref{SSmasterderive} the complete master equation resembles
above equation only in the limit of no squeezing $\xi\to 0$ -- see
Eq.~(\ref{Eredmaster}) -- and even then with a time-dependent
damping constant $\gamma_D$.
\end{itemize}

Instead, the complete master equation (\ref{Emaster}) 
displays more similarities to
one in a squeezed thermal bath where one has to replace the parameters
by time-dependent functions. Accordingly, the dynamical system under
consideration is described properly only by an explicitly
time-dependent master equation.

Potential limitations to Eq.~(\ref{Edodonov}) have already
been anticipated in \cite{leakydodonov}.


\section{The Non-Perturbative Approach}\label{Snonperturb}

The previous results in Sec.~\ref{SScreation1} and
Sec.~\ref{SScreation2} have not been able to explain the behavior of
the system in the limit of a long-lasting disturbance. 
The leading order term $T {\cal S} (2 T)$ in (\ref{Equadleftresp}) has
a negative sign which would lead to negative particle numbers for
large disturbance times $T$. This problem can only be solved by
including all orders in $\eta$. In this section we present a
non-perturbative approach within the RWA which enables a convenient 
calculation of expectation values using computer algebra systems
\cite{schaller}. 
As a further advantage we want to mention that it can in principle 
be generalized in a straightforward way to the case of more than just 
two coupling modes as was assumed in Sec.~\ref{SSrwa}.


\subsection{Time Evolution}\label{SSevolution}

Application of the RWA in Sec.~\ref{SSrwa} yielded the effective 
time-evolution operator $\hat{U}_{\rm eff}=\exp(-i\hat{H}_{\rm eff}^I T)$
with the effective interaction Hamiltonian (\ref{Eheff}). 
We want to calculate the expectation value of an observable $\hat{Y}$
\bea 
  \expval{Y(T)} & = & \trace{\hat{Y} e^{-i \hat{H}_{\rm eff}^I T} \hat{\rho}_0 
                e^{+i \hat{H}_{\rm eff}^I T}}\nn\\ 
             & = & \trace{e^{+i \hat{H}_{\rm eff}^I T} \hat{Y} 
                e^{-i \hat{H}_{\rm eff}^I T}\hat{\rho}_0}\,.
\eea
In contrast to the previous sections here the full time dependence 
is shifted back on the operator $\hat{Y}$. Since $\hat{Y}$ can be 
expressed using creation and annihilation operators and due to 
unitarity of the time-evolution operator one just has to find a
solution for the full time dependence of the ladder operators which
is given by
\bea
  \hat{a}_\sigma(T) = e^{+i \hat{H}_{\rm eff}^I T} \hat{a}_\sigma 
  e^{-i \hat{H}_{\rm eff}^I T}\,.
\eea
The above expression requires special care to evaluate, since
$\hat{H}_{\rm eff}^I$ is not a pure squeezing generator. 
In our considerations the effective interaction Hamiltonian is 
time-independent which does not necessarily hold in general. To
preserve generality we will therefore introduce an auxiliary 
parameter $\vartheta$ while keeping the time $T$ fixed. This 
enables us to write 
\bea
  \hat{a}_\sigma(\vartheta) = e^{+i \hat{H}_{\rm eff}^I T \vartheta} 
                \hat{a}_\sigma  e^{-i \hat{H}_{\rm eff}^I T \vartheta}\,.
\eea
Obviously we are interested in 
$\hat{a}_\sigma(T)=\hat{a}_\sigma(\vartheta=1)$. To this end we define 
a 4 dimensional column vector (see also \cite{dalvit})
\bea
  \underline{\hat{x}} (\vartheta) = \left( \begin{array}{c} 
                \hat{a}_L (\vartheta) \\  
                \hat{a}_L^\dagger (\vartheta)\\ 
                \hat{a}_R (\vartheta)\\
                \hat{a}_R^\dagger (\vartheta) 
                \end{array} \right)\,.
\eea
Since $\hat{H}_{\rm eff}^I$ does not depend on $\vartheta$, one finds
\bea
  \tdiff{\underline{\hat{x}}}{\vartheta} = 
  i T \commute{\hat{H}_{\rm eff}}{\underline{\hat{x}}(\vartheta)} 
                = T\underline{A}\,\underline{\hat{x}}(\vartheta) 
\eea
where $\underline{A}$ is a number-valued 4 by 4 matrix acting on 
$\underline{x}$. This form can always be reached if the effective 
Hamiltonian is quadratic: The commutation relations
(\ref{Ecommrel}) lead to a linear combination of creation and annihilation
operators that can always be written as a number-valued matrix 
$\underline{A}$ acting on $\underline{x}$.
Since $\underline{A}$ is independent of $\vartheta$ the solution is
obtained via
\bea 
  \underline{\hat{x}}(\vartheta) = \exp\left(\underline{A} T \vartheta\right) 
                \underline{\hat{x}}(0)
\eea
and hence
\bea
  \underline{\hat{x}}(1) = \exp\left(\underline{A} T \right) 
                \underline{\hat{x}}(0)
                =\underline{U}(T)\underline{\hat{x}}(0)\,.
\eea
Thus the whole problem reduces to a calculation of the time-evolution matrix
$\underline{U}(T)=\exp(\underline{A} T)$. In the present case
the structure of $\hat{H}_{\rm eff}$ in Eq.~(\ref{Eheff}) implies
a very simple form of $\underline{A}$ 
\bea
  {\underline{A}}  = \left( \begin{array}{cccc} 
                        0 & 2\xi & \chi & 0\\ 
                        2\xi & 0 & 0 & \chi\\
                        -\chi & 0 & 0 & 0\\
                        0 & -\chi & 0 & 0 
                     \end{array} \right)\,.
\eea
The four eigenvalues of $\underline{A}$ are given by
\bea\label{Eeigenvalues}
  \lambda_1 & = & \xi+\sqrt{\xi^2-\chi^2}\,,\nn\\ 
  \lambda_2 & = & \xi-\sqrt{\xi^2-\chi^2}\,,\nn\\ 
  \lambda_3 & = & -\xi+\sqrt{\xi^2-\chi^2}\,,\nn\\ 
  \lambda_4 & = & -\xi-\sqrt{\xi^2-\chi^2}\,. 
\eea
For reasons of brevity we shall omit the full listing of the matrix 
$\underline{U}(T)=\exp(\underline{A} T)$ -- it can easily be calculated
using some computer algebra system. 
Note that the exponential matrix $\underline{U}(T)$ is positive definite for
all $T$ and thus will not exhibit the problems associated with the 
extrapolation of the used approximations of sections \ref{Squadresp} and 
\ref{Smaster} beyond their range of validity.
In order to calculate expectation values one just needs the
matrix elements of $\underline{U}(T)$. This becomes
evident considering the time evolution of the new annihilation
and creation operators 
$\hat{\underline{x}} (T) = \underline{U}\,(T) \hat{\underline{x}} (0)$, i.e. 
$\hat{x}_i (T) = \sum_{j=1}^4 U_{ij} (T) \hat{x}_j (0)$.
Therefore the expectation values of particle number operators 
of the resonance modes 
$\hat{N}_{L}(T)={\hat{x}}_2(T){\hat{x}}_1(T)$ and 
$\hat{N}_{R}(T)={\hat{x}}_4(T){\hat{x}}_3(T)$ can be 
calculated simply by insertion of
$\hat{\underline{x}} (T)$. After evaluation of the remaining traces
containing only the initial creation and annihilation operators
${\hat{x}}_i (0)$ one finds
the full response function to be a combination of matrix 
elements of $\underline{U}(T)$
\bea \label{Efulln}
  \expval{N_{L}(T)} & = & (U_{12}U_{21} + U_{14}U_{23})\nonumber\\ 
                 &   & +(U_{11}U_{22} + U_{12}U_{21}) \expval{N_{L}^0}
                        \nonumber\\ 
                 &   & +(U_{13}U_{24} + U_{14}U_{23})\expval{N_R^0}
\eea
and
\bea
  \expval{N_{R}(T)} & = & (U_{41}U_{32} + U_{34}U_{43})\nonumber\\ 
                 &   & +(U_{42}U_{31} + U_{41}U_{32}) \expval{N_{L}^0}
                        \nonumber\\ 
                 &   & +(U_{33}U_{44} + U_{43}U_{34})\expval{N_R^0}\,.
\eea


\subsection{Particle Creation}\label{SScreation3}

\begin{figure}
  \includegraphics[width=9cm]{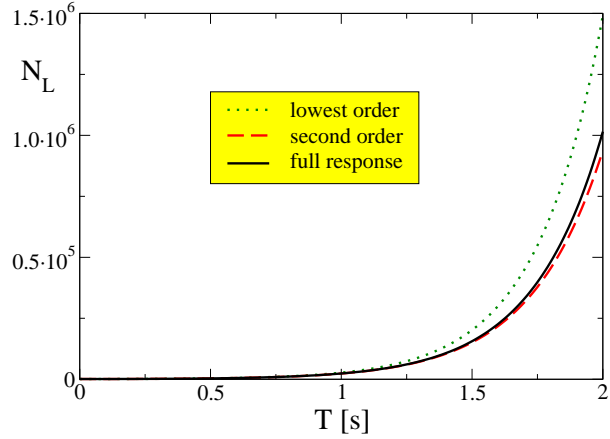}
  \caption{\label{Fanswercompleft}
  Particle creation in the fundamental resonance mode $N_{L}$ for
  $N_L^0=1000$, $N_R^0=100$, $\xi=1$ Hz, and $\chi=0.5$ Hz. An exponential
  growth is found in all cases.} 
\end{figure}

With the full knowledge of $\underline{U}$ we are now in a 
position to state the full response function of the particle number operator. 
Having inserted the matrix elements of $\underline{U}$ 
into Eq.~(\ref{Efulln}) one finds after performing some simplifications 
\bea\label{Enleft}
  \expval{N_{L}(T)} & = & \frac{1}{4
        \left(\xi^2-\chi^2\right)}\times\nn\\ 
  &   & \left\{ 
  \xi\cosh\left[2T\left(\xi+\sqrt{\xi^2-\chi^2}\right)\right]  
  \left[\xi+\sqrt{\xi^2-\chi^2}\right]\right.\nn\\ 
  &   &  
  +\xi\cosh\left[2T\left(\xi-\sqrt{\xi^2-\chi^2}\right)\right]  
  \left[\xi-\sqrt{\xi^2-\chi^2}\right]\nn\\ 
  &   &  
  \left.-2\chi^2\cosh\left[2T\xi\right] 
  \stackrel{\phantom{m}}{-} 2\left(\xi^2-\chi^2\right)\right\}\nn\\ 
  \nn\\ 
  &   & + \frac{N_{L}^0}{2 (\xi^2-\chi^2)}\times\nn\\ 
  &   & \left\{ 
  \xi \cosh\left[2T\left(\xi+\sqrt{\xi^2-\chi^2}\right)\right]  
  \left[\xi+\sqrt{\xi^2-\chi^2}\right]\right.\nn\\ 
  &   &  
  +\xi \cosh\left[2T\left(\xi-\sqrt{\xi^2-\chi^2}\right)\right]  
  \left[\xi-\sqrt{\xi^2-\chi^2}\right]\nn\\ 
  &   &  
  \left.-\chi^2\cosh\left[2T\xi\right] 
  \left[\cosh\left(2T\sqrt{\xi^2-\chi^2}\right)+1\right] 
        \right\}\nn\\ 
  \nn\\  
  &   & +\frac{N_R^0}{2(\xi^2-\chi^2)}\times\nn\\  
  &   & \left\{ 
        \chi^2\cosh\left[2T\xi\right] 
        \left[\cosh\left(2T\sqrt{\xi^2-\chi^2}\right)-1\right] 
        \right\}\,.\nn\\
\eea 
This result is valid to all orders in $\chi/\xi$ or $\eta$,
respectively. To show consistency with 
the results obtained in Sec.~\ref{SScreation1} and Sec.
\ref{SScreation2} we expanded the above expression around $\chi/\xi=0$ up
to second order and found complete agreement with Eq.~(\ref{Equadleftresp})!
However, even for large values of $\chi/\xi=1/2$ the quadratic
approximation is a rather good one -- provided that the duration of the
disturbance $T$ is not extremely large -- as one can see in 
Fig.~\ref{Fanswercompleft}.\\

\begin{figure}
  \includegraphics[width=9cm]{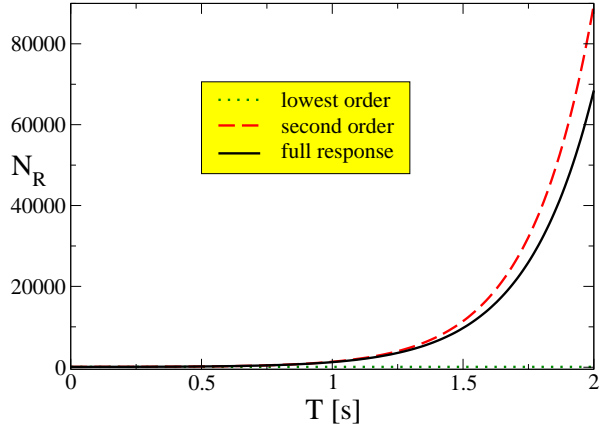}
  \caption{\label{Fanswercompright}
  Particle creation in the right resonance mode $N_{R}$ 
  for $N_L^0=1000$, $N_R^0=100$, $\xi=1$ Hz, and $\chi=0.5$ Hz. The lowest
  order result just corresponds to a constant initial particle
  number. Generally the particle creation in the reservoir is a much
  smaller effect then in the leaky system, see also
  Fig.~\ref{Fanswercompleft}. } 
\end{figure}

Doing the same calculations for the corresponding right-dominated mode 
one finds as a result
\bea\label{Enright}
  \expval{N_{R}(T)} & = & \frac{1}{4 (\xi^2-\chi^2)}\times\nn\\ 
  &   & \left\{ 
  \xi \cosh\left[2T\left(\xi+\sqrt{\xi^2-\chi^2}\right)\right]  
  \left[ \xi-\sqrt{\xi^2-\chi^2}\right]\right.\nn\\ 
  &   &  
  +\xi\cosh\left[2T\left(\xi-\sqrt{\xi^2-\chi^2}\right)\right]  
  \left[\xi+\sqrt{\xi^2-\chi^2}\right]\nn\\ 
  &   &  
  \left.-2\chi^2\cosh\left[2T\xi\right] 
  \stackrel{\phantom{m}}{-} 2\left(\xi^2-\chi^2\right)\right\}\nn\\ 
  \nn\\  
  &   & +\frac{N_L^0}{2(\xi^2-\chi^2)}\times\nn\\  
  &   & \left\{ 
        \chi^2\cosh\left[2T\xi\right] 
        \left[\cosh\left(2T\sqrt{\xi^2-\chi^2}\right)-1\right] 
        \right\} \nn\\   
  \nn\\ 
  &   & + \frac{N_R^0}{2 (\xi^2-\chi^2)}\times\nn\\ 
  &   & \left\{ 
  \xi \cosh\left[2T\left(\xi+\sqrt{\xi^2-\chi^2}\right)\right]  
  \left[\xi-\sqrt{\xi^2-\chi^2}\right]\right.\nn\\ 
  &   &
  +\xi \cosh\left[2T\left(\xi-\sqrt{\xi^2-\chi^2}\right)\right]  
  \left[\xi+\sqrt{\xi^2-\chi^2}\right]\nn\\ 
  &   &  
  \left.-\chi^2\cosh\left[2T\xi\right] 
  \left[\cosh\left(2T\sqrt{\xi^2-\chi^2}\right)+1\right] 
        \right\} \nn\\
\eea
where the remarkable agreement of coefficients of $N_R^0$ in
$\expval{N_L}$ and of $N_L^0$ in $\expval{N_R}$ as was already noticed
in Sec.~\ref{SScreation1} is preserved for all orders in $\eta$. 
These terms fit the classical picture of particle transportation 
through the leaky membrane where the particle flux is proportional 
to the number of particles on the other side.
Again, expanding around $\chi/\xi=0$ up to second order yields exact
agreement with (\ref{Equadrightresp}).
Accordingly, also outside the leaky cavity
particles are produced due to the dynamical Casimir effect, see also
Fig.~\ref{Fanswercompright}.\\
Note that at least the quadratic answer is necessary to treat particle
creation effects outside the leaky cavity.
It is still valid that finite-temperature corrections will enhance the 
pure vacuum phenomenon of particle production by several orders of 
magnitude. (For a direct
comparison see Fig. \ref{Fanswercomptemp} in Sec.~\ref{Sdisc}.)


\subsection{Further Remarks}\label{SSremarks}

We have derived a complete solution for the effective interaction
Hamiltonian (\ref{Eheff}) which is valid to all orders in
$\chi/\xi=\order{\eta}$. 
As an illustration we consider a case outside our initial intentions
where $\chi/\xi$ also assumes large values, e.g. $\chi/\xi \ge 1$.
In this case the arguments of the hyperbolic functions in
(\ref{Enleft}) and (\ref{Enright}) will receive an 
imaginary part. The arising imaginary parts of
$\expval{N(T)}$ cancel as they have to because $\hat{N}$ is a physical
observable. Thus one finds that if the velocity parameter $\chi$ exceeds 
the critical value
$\chi\ge\xi$ the particle occupation number of the resonance modes
versus the vibration time will exhibit oscillations. Of course, for
the case of a nearly perfectly reflecting mirror inside this scenario is
completely unrealistic since then $\chi/\xi=\order{\eta}$ will be 
relatively small. 
However, this case is not at all academic: If the label $R$
stood for a left-dominated mode -- which is the case we excluded
in our considerations so far and whose equivalent for ideal cavities
has been considered in \cite{dalvit} -- $\chi/\xi$ may very well
become large, since $m_{LR}$ would then be of $\order{1}$. 
\begin{figure}
  \includegraphics[width=9cm]{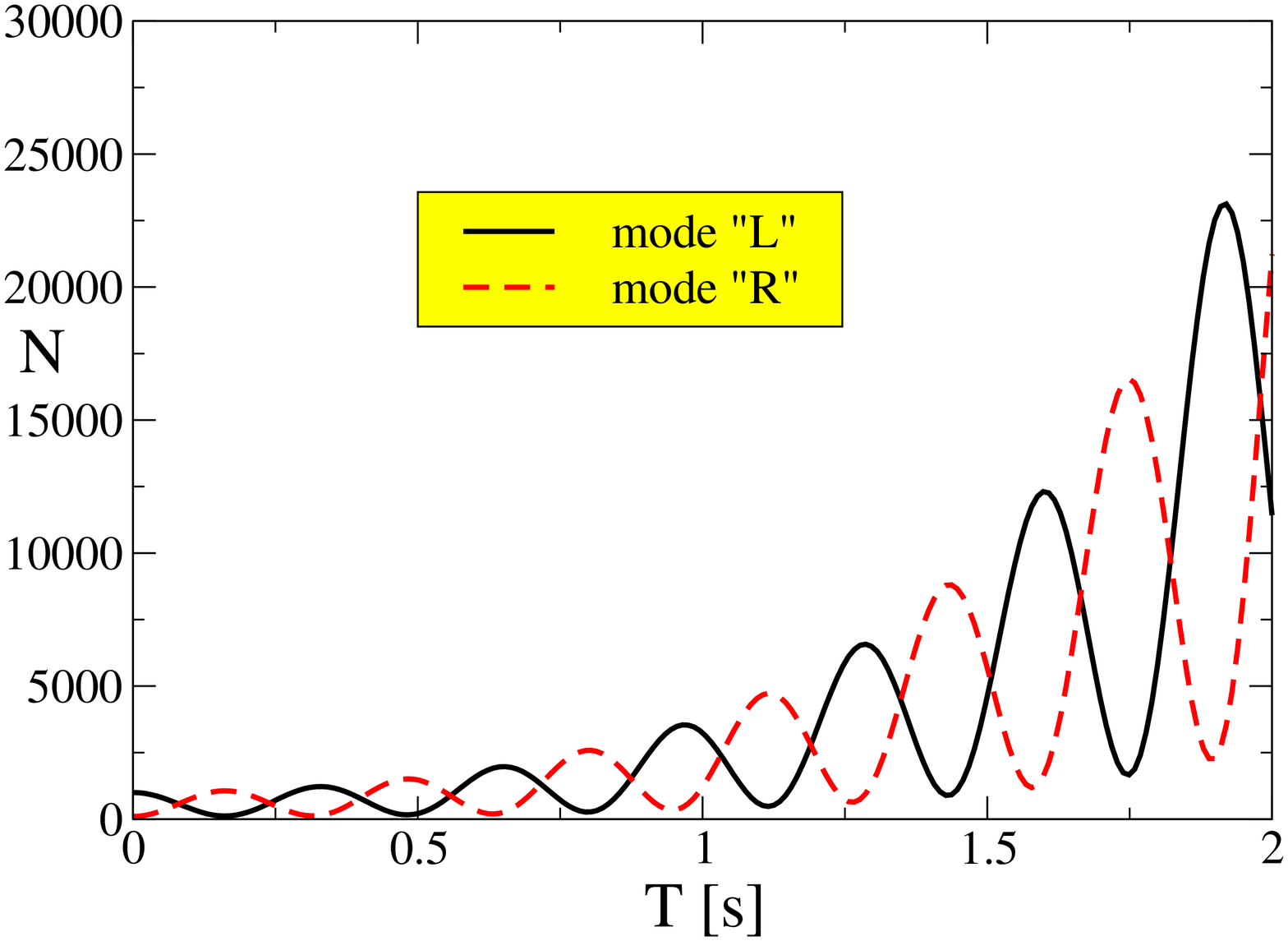}
  \caption{\label{Foscillation}
  Oscillations of the particle number in the  
  resonance modes $N_L$ and $N_R$ for
  $N_L^0=1000$, $N_R^0=100$, $\xi=1$ Hz and $\chi=11$ Hz.} 
\end{figure}

Similar oscillations of the particle number were also found in the 
case of strong inter-mode coupling in an ideal cavity \cite{dalvit}.

Note that with $\chi/\xi=\order{\eta}$ this also leads to an
upper bound for the internal mirror transmittance $\eta$ above which
(corresponding to a highly transparent mirror) one finds oscillations 
that correspond to inter-mode coupling rather than to system-reservoir
coupling. From another perspective this phenomenon could also be
envisaged as follows: Starting with an ideal cavity whose original
dimensions do not permit inter-mode coupling one can insert a highly
transparent mirror ($\eta \gg 1$). This mirror in turn detunes the 
ideal cavity in such a way that it now permits inter-mode couplings 
as well.

It is remarkable that in Fig.~\ref{Foscillation} the phase of the two
modes is shifted: When $\expval{N_L(T)}$ is at its maximum, then
$\expval{N_R(T)}$ is at its minimum and vice versa. This fits nicely
with the picture of mode hopping mediated by the inter-mode coupling
$\chi$. One even observes a decrease of the particle number in the
$L$-mode for small times. When defining an effective temperature
\cite{finitetemp} this would correspond to an effective cooling of 
the $L$-mode.
An extreme case of this consideration would be the limit of no
squeezing, i.e. $\xi = 0$. This would correspond to the possibility
(see also Sec.~\ref{SSrwa}) of not fulfilling the squeezing but the
velocity resonance condition. Performing the limit $\xi \to 0$
everywhere in equations (\ref{Enleft}) and (\ref{Enright}) one would
find pure oscillations of the particle numbers and no exponential
growth at all. This case is therefore counterproductive for an
experimental verification. Note however, that this is different for
the case of $\oplus$-coupling.
The consistency with the earlier results leads to the conclusion that
our approach was justified and the full response function should
describe the rate of particle production correctly within the RWA.

Please note that the described procedure also holds for more
than just two coupling modes: if one has e.g. $n$ modes fulfilling
the resonance conditions given in Sec.~\ref{SSrwa}, the formalism still holds
and one will have to define a $2n$ dimensional vector $\underline{x}$.
Of course then $n$ creation and $n$ annihilation operators of these
resonance modes will be contained in the
Hamiltonian and therefore also $\underline{A}$ as well as
$\underline{U}$ will be $2n$ by $2n$ matrices. 
The calculations will simply become more involved but can certainly be 
performed, e.g. by means of computer algebra systems.


\section{Detuning}\label{Sdetune}

So far we did assume an exact fulfillment of the resonance
conditions, i.e. the vibration of the left cavity wall did match
exactly twice the fundamental resonance frequency $\Omega^0_L$. 
However, in real situations one will of course have to deal with deviations
from this desired external frequency since it will not be possible 
to match it with arbitrary precision. 
In addition, the back-reaction of the created quanta might cause 
the external vibration frequency to change.

Consequently, we will now discuss the detuned situation -- where $\omega$
assumes slightly off-resonant values. For a review
of detuning effects see e.g. \cite{leakydodonov,dalvit,dodonovdetuned}.
It has been shown in the literature that there exist threshold values
for the detuning, above which the exponential creation of particles disappears. 

Unfortunately the RWA used in our previous considerations cannot 
simply be generalized to this situation. For a slight deviation from
the resonance conditions in equations (\ref{Erescond1}) and
(\ref{Eresvelocity}) the terms with time-ordering -- see section 
\ref{SSrwa} -- are no longer negligible in this way.

We will consider slightly off-resonant situations, where the external 
vibration frequency does not match the fundamental resonance exactly
\bea
  \omega = 2\Omega_L^0(1+\delta)\,,
\eea
where $\delta$ denotes a small (dimensionless) deviation $\delta\ll1$.  
However, if one considers such a variance it is only consequent to
include a possible discrepancy of the coupling resonance as well, 
cf.\ \cite{dodonovdetuned} 
\bea\label{Eadddeviation}
  \Omega_R^0 = \Omega_L^0(3+\Delta)\,,
\eea
where $\Delta\ll1$ denotes the deviation of the coupling right-dominated 
mode from the $\ominus$-coupling resonance condition with the fundamental 
resonance mode. 
We will adapt the multiple scale analysis (MSA) as proposed in 
\cite{dalvit} to our scenario of a leaky cavity, see also 
\cite{dodonovdetuned}. 
For this purpose we restrict to the results, since the steps in 
\cite{dalvit,dodonovdetuned} can strictly be followed -- see also 
appendix \ref{Sappendix}.
The main difference in these considerations is that we use the eigenfunction 
system of subsection \ref{SSeigenmodes} instead of those of an ideal cavity 
and that we assume the additional deviation (\ref{Eadddeviation}) -- see also 
\cite{dodonovdetuned}.

In analogy to subsection \ref{SSevolution} one obtains a matrix 
${\underline{A}'}$ governing the time evolution of the ladder operators
\bea
\underline{\hat{x}}'=\left(\begin{array}{c}
    \hat{a}_L\\
    \hat{a}_L^\dagger\\
    \hat{a}_R/\sqrt{3}\\
    \hat{a}_R^\dagger/\sqrt{3}
  \end{array}\right)\,.
\eea
The creation of quanta will only be exponential -- and thus noticeable
-- if at least one of the eigenvalues of this matrix
\beann
  {\underline{A}'} \approx  
  \left( \begin{array}{cccc}
      i\Omega_L^0\delta & 2\xi & \sqrt{3}\chi & 0 \\
      2\xi & -i\Omega_L^0\delta & 0 & \sqrt{3}\chi \\
      -\chi/\sqrt{3} & 0 & i\Omega_L^0(3\delta-\Delta)  & 0 \\
      0 & -\chi/\sqrt{3} & 0 & -i\Omega_L^0(3\delta-\Delta)
    \end{array}\right)\,,
\eeann
does have a positive real part. 
Note that the slight disagreement between the above matrix and the one given
in Ref.\ \cite{dodonovdetuned} is caused by the usage of a different phase 
($\sin$ instead of $\cos$).
With the abbreviations
\beann
  {\cal U}=8\xi^2-4\chi^2+12(\Omega_L^0)^2\delta\Delta
  -2(\Omega_L^0)^2\Delta^2
  -20(\Omega_L^0)^2\delta^2\,,
\eeann
and
\bea
  {\cal V}&=&
  +16\xi^2(\xi^2-\chi^2)\nn\\
  &&+64(\Omega_L^0)^2\delta^2[\xi^2+\chi^2+(\Omega_L^0)^2\delta^2]\nn\\
  &&+(\Omega_L^0)^2\Delta^2[8\xi^2+4\chi^2 
  +52(\Omega_L^0)^2\delta^2+(\Omega_L^0)^2\Delta^2]\nn\\
  &&-4(\Omega_L^0)^2\delta\Delta[12\xi^2+8\chi^2+
  24(\Omega_L^0)^2\delta^2+3(\Omega_L^0)^2\Delta^2]\,,\nn\\
\eea
the eigenvalues of the above matrix read 
(cf.\ \cite{dalvit,dodonovdetuned})
\bea
\lambda=\pm\frac{1}{2}\sqrt{{\cal U}\pm2\sqrt{\cal V}}
\,.
\eea
As a consistency check we may set $\delta=\Delta=0$ where the eigenvalues 
reduce to the ones given in Eq.\ (\ref{Eeigenvalues}).
On the other hand, for $\chi=0$ one recovers the usual result of pure
squeezing in an ideal cavity $\lambda_+=\pm\sqrt{4\xi^4-\Omega_L^2\delta^2}$. 

Note that in contrast to \cite{dalvit,dodonovdetuned} the inter-mode coupling
and thus the parameter $\chi$ is very small $\chi\ll\xi$.
This enables us to expand the quantities $\cal U$ and $\cal V$ into powers of 
$\chi$.
The condition for a real eigenvalue ${\cal U}+2\sqrt{\cal V}>0$ reads
\bea
  \Omega_L^2\delta^2 &<& 
  4\xi^2
  -2\frac{4\xi^2+\Omega_L^2\delta\Delta-4\Omega_L^2\delta^2}
  {4\xi^2-\Omega_L^2\delta^2+\Omega_L^2(3\delta-\Delta)^2}
  \,\chi^2\nn\\
  &&+\order{\chi^4}\,.
\eea
Since $\chi$ is supposed to be small $\chi\ll\xi$ one obtains a significant
contribution only if $3\delta\approx\Delta$ and also in this case merely
in the immediate vicinity of the critical value $2\xi=\Omega_L^0\delta$.
Consequently, the presence of an internal mirror of moderate quality do not 
drastically modify the threshold 
\bea
\delta < \frac{1}{2}\left(\frac{\Omega_L^{x}}{\Omega_L^0}\right)^2 \epsilon
\eea
for exponential particle creation. 
However, we would like to emphasize that the shifts of the eigenfrequencies 
of the cavity due to the partly permeable internal wall must be taken into 
account, see also section \ref{Sconclude} below.


\section{Summary}\label{Ssum}

We have considered a massless scalar quantum field inside a leaky
cavity modeled by means of a dispersive mirror. For the case of the lossy
cavity vibrating at twice the fundamental resonance frequency we
derived an effective Hamiltonian using the rotating wave approximation.
Within the framework of response theory the magnitude of particle
creation due to the dynamical Casimir effect was calculated.
Furthermore we deduced the corresponding master equation via applying
the Born-Markov approximation. We found a discrepancy to the master
equations used so far (see \cite{leakydodonov}) to describe 
oscillating leaky cavities.
We also applied a non-perturbative approach for the explicit
calculation of the time evolution starting from the effective
Hamiltonian. 
All these methods were found to lead to consistent results.
In addition, the effects of a detuned external vibration frequency
need to be taken into account.

It turned out that for the case of moderately low transmission
coefficients (or sufficient quality factors) the rate of created
particles is almost the same as for ideal cavities.
The squeezing of the fundamental resonance mode as well as the strong
enhancement of particle production due to the dynamical Casimir effect
are preserved in the presence of transparent mirrors.


\section{Discussion}\label{Sdisc}

In order to illustrate reasonable magnitudes let us specify the
relevant parameters:
A cavity with a typical size of $\Lambda=1$ cm would have a
fundamental resonance frequency of $\Omega_{L}^0\approx 150$ GHz
i.e., the corresponding coupling right dominated mode must have a
frequency of about $\Omega_{R}^0=3\Omega_{L}^0\approx 450$ GHz. 
According to Ref. \cite{leakydodonov} we assume a dimensionless 
vibration amplitude $\epsilon \approx 10^{-8}$.
Consequently, in order to create a significant amount of particles one
would have to sustain the external oscillations over an interval of
several milliseconds.
At room temperature $1/\beta \approx 300$ K one finds the initial
particle occupation numbers to be $N_L^0 \approx 240$ and 
$N_R^0 \approx 80$. Using the above values the squeezing parameter
determines to $\xi\approx 150\;{\rm Hz}$. 

As the quality factor $Q$ of a resonator is defined as \cite{jackson}
\bea
  Q = 2 \pi \frac{{\rm energy\;in\;cavity}}
        {{\rm energy\;loss\;per\;period}}\,,
\eea
one finds as a classical estimate yields for our system 
\bea
  Q = \frac{2 \pi}{{\mathcal{\abs{T}}}^2} 
        = 2 \pi \left(1+\left(\frac{\gamma}{\Omega_L^x}\right)^2 \right) 
        = \order{\frac{1}{\eta^2}} \,.
\eea
${\mathcal{T}}$ denotes the transmission amplitude through the internal
dispersive mirror and $\gamma$ was defined in Sec.~\ref{Sbasic}.  
Assuming a $Q$-factor of $Q\approx 10^8$ \cite{leakydodonov} (and
references therein) this would imply for the corresponding
perturbation parameter $\eta=\Omega_L^x/\gamma = \order{10^{-4}}$. 
With these values, a reasonable velocity parameter could be given by 
$\chi\approx 2\;{\rm mHz}$. 

\begin{figure}
  \includegraphics[width=9cm]{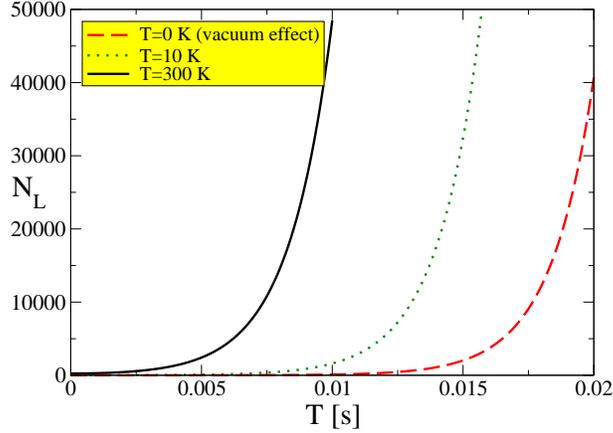}
  \caption{\label{Fanswercomptemp}
  Comparison of particle production in the fundamental resonance mode 
  at finite temperature and without temperature effects. At room
  temperature (300 K) the initial occupation numbers result as 
  $N_L^0=240$ and $N_R^0=80$. Accordingly, squeezing and velocity
  parameters are given by $\xi=150\;{\rm Hz}$ and $\chi=2 \;{\rm
  mHz}$. At room temperature the particle number
  reaches significant values much faster.} 
\end{figure}

The particle content of the leaky cavity is depicted in 
Fig.~\ref{Fanswercomptemp}.\\


\section{Conclusion}\label{Sconclude}

\begin{figure}
  \includegraphics[width=8cm]{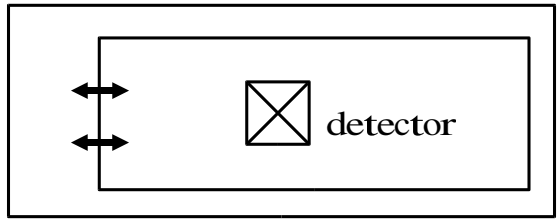}
  \caption{\label{Fproposal}
  Sketch of a vibrating cavity enclosed by a larger one. This
  configuration may facilitate the experimental verification of the
  dynamical Casimir effect inside the smaller cavity.} 
\end{figure}

According to the above considerations it is necessary to vibrate 
several milliseconds in order to produce measurable effects. 
As already stated, a cavity at finite temperature might even 
be advantageous -- provided the cavity is still nearly ideal at 
its characteristic thermal wavelength.
However, even after only one millisecond ($10^8$ periods) a classical estimate
based on a quality factor of $Q = 10^8$ would indicate drastic energy 
losses. On the other hand, our calculations based on a complete quantum
treatment show that the effects of losses are almost negligible
compared to the rate of particle creation as long as $\eta \ll 1$. 
This leads to the conclusion that lower cavity quality factors than
proposed in \cite{leakydodonov}, e.g. $Q=10^6$ 
[implying $\eta=\order{10^{-3}}$] would already completely suffice to 
justify our approximations \cite{schaller}. 
Such quality factors are within the reach of the
current experimental status. Of course our calculations
are based on the assumption that the larger cavity -- including both the 
reservoir and the leaky cavity -- is perfectly conducting. The error
made by this presumption is of $\order{Q^{-2}}$ and therefore
certainly negligible. 
Consequently, the experimental verification of the dynamical Casimir
effect could be facilitated by a configuration where the vibrating
cavity is enclosed by a slightly larger one as is demonstrated in 
Fig.~\ref{Fproposal}.
A further important result \cite{schaller} of our investigations is the 
shift of the cavity eigenfrequencies (\ref{Eeigenfreq}) of
$\order{\eta}=\order{10^{-3}}$ which needs to be taken into account in
order to make an experimental observation of quantum radiation
feasible. 


\section{Outlook}\label{Sout}


\subsection{Multi-Mode Coupling}\label{SSmultimode}

In section \ref{SSrwa} we assumed that only one of the right-dominated
modes fulfills the resonance condition for the velocity
Hamiltonian, i.e., exactly two modes are coupled. If, however, the 
reservoir cavity becomes larger, the spacing between different levels
of its spectrum decreases so that eventually more than just one
right-dominated mode begin to couple -- at least within the range of
detuning. In this case the effective velocity Hamiltonian would
constitute a sum of single two-mode coupling Hamiltonians as the one
in Eq.~(\ref{Eeffhv}) -- but accounting for different right modes
$R_1$, $R_2$, etc. As we have observed in Fig.~\ref{Fanswercompleft}
in Sec.~\ref{SScreation3}, the quadratic answer is completely
sufficient for reasonable values of $\chi/\xi$. Inserting the
aforementioned sum of
Hamiltonians into the quadratic answer one observes that the mixing
terms vanish. Since the
effective velocity Hamiltonian only contains odd powers of the
creation and annihilation operators per mode, one can only obtain a
non-vanishing trace if it involves two operators of the same (right) 
mode. 
Therefore the quadratic answer also decomposes into a
sum of contributions each accounting for one right-dominated mode. 
Hence we expect the general structure of $\expval{N_L(T)}$ 
to persist -- just substitute
$\chi^2 \to \sum_i \chi_i^2$ in the leading 
contributions\footnote{Of course, multi-mode coupling can also be
taken care of by the non-perturbative approach of
Sec.~\ref{Snonperturb}. The required effort, however, will increase very
fast, since the corresponding matrix grows with the number of coupling 
modes.}. 
In order to ensure the applicability of the perturbative treatment the
number $n$ of coupling right-dominated modes has to be small enough to
satisfy $\sum_i \chi_i^2/\xi^2 \approx n \chi^2/\xi^2 \ll 1$. 

Typically the spacing between two neighboring right-dominated modes is
of $\order{1/L}$, i.e., the inverse of the characteristic length of
the reservoir. On the other hand, the width of the resonance peak is of
$\order{1/T}$. Consequently, if the duration of the disturbance $T$ 
(1 ms) exceeds the characteristic length of the reservoir (which is
the case for $L<10^5$ m) then $n$ is certainly small enough. 


\subsection{Electromagnetic Field}\label{SSemfield}

So far, we have considered a noninteracting, massless, and neutral 
scalar field. The next step could be to extend the calculations to the
electromagnetic field. In this case several new difficulties arise:

\begin{enumerate}
\item
The boundary conditions cannot just simply be described by Dirichlet
(or Neumann) conditions. Especially for moving walls their form will
be more complicated due to Ampere's law (mixing of {\boldmath{$E$}} and
{\boldmath{$B$}}). 
\item
As the electromagnetic field is a gauge theory, one has to eliminate
the unphysical degrees of freedom in order to quantize it. Again, for
dynamic external conditions this requires special care, see
e.g. \cite{quantumrad}.
\item
The different polarizations of photons need to be taken into account
which are of special interest concerning the fulfillment of the
resonance conditions. 
\end{enumerate}
According to \cite{jackson} the eigenmodes of the stationary cavity can
be divided into  TE and TM modes. For several cavities
(rectangular, cylindrical, spherical) the eigenfrequencies are
well-known. This enables one to determine the squeezing part of the
interaction Hamiltonian. 

In order to deduce the velocity Hamiltonian it will be necessary to
find an appropriate model for the dispersive mirror. This can be
achieved by using a thin dielectric slab with a high
permittivity: $\varepsilon (x) = 1 + \gamma \delta (x)$. As has been
shown for a stationary system in \cite{scully} this leads to a similar 
eigenvalue equation as (\ref{Eeigenvalue}).\\

For the detection of the created field quanta some detecting device
will have to be placed inside the cavity, e.g. an atom. However, the detector
will always influence the created field as well. A simple approach for the 
modeling of a two-level system has been provided in \cite{dodonov3,dodonov5}. 
In addition, the non-adiabatic parametric modulation  of the atomic Lamb 
shift -- as has been considered in \cite{lambversuscasimir} -- must be taken 
into account, since it will cause excitations of the atom as well.

Note that the induced quantum field will also excite the internal 
degrees of freedom of the cavity mirrors -- an alternate description of 
losses should therefore also take the energy dissipation of the losses 
within the mirrors into account, see e.g. \cite{leakylaw}.

Future work combining all these effects is of immense importance regarding
experiments on quantum radiation using the dynamical Casimir effect.


\section{Acknowledgments}\label{Sthanks}

The authors are indebted to A.~Calogeracos for fruitful discussions. 
R.~S.~acknowledges financial support by the Alexander von Humboldt
foundation and NSERC. 
G.~P.~acknowledges financial support by BMBF, DFG, and GSI.


\begin{appendix}
\section{Multiple Scale Analysis}\label{Sappendix}

Starting with the Lagrangian (\ref{lagrangian}) it is straightforward 
to show that the field operator fulfills a modified wave equation
\bea
  \left\{ \Box + 2 V(\mbox{\boldmath $r$};t) \right\} 
        \hat{\varphi}(\mbox{\boldmath $r$},t) = 0\,.
\eea
If one now follows \cite{dalvit} by introducing ladder operators via
the expansion
\bea
  \hat{\varphi}(\mbox{\boldmath $r$},t) = \sum_n \hat{a}_n^{\rm in}
        u_n (\mbox{\boldmath $r$},t) + {\rm h.c.}\,,
\eea
where
\bea
  u_n (\mbox{\boldmath $r$},t<0) & = & \frac{1}{\sqrt{2\Omega_n^0}} 
        f_n (\mbox{\boldmath $r$}) e^{-i\Omega_n^0 t}\,,\\
  u_n (\mbox{\boldmath $r$},t>0) & = & \sum_k Q_k^{(n)} (t) 
        f_k (\mbox{\boldmath $r$};t)\,,
\eea
one can derive a time evolution equation for the coefficients
$Q_k^n(t)$. Using the properties (\ref{Efeatures}) of the eigenfunctions 
one obtains
\bea\label{Edalvit1}
  \ddot{Q}_m^{(n)}(t)+\Omega_m^2(t) Q_m^{(n)}(t)
        & = & 2 \lambda(t) \sum_k g_{mk} \dot{Q}_k^{(n)}(t)\nn\\
        && + \dot{\lambda}(t) \sum_k g_{mk} \dot{Q}_k^{(n)}(t)\nn\\
        && + \lambda^2 (t) \sum_{k,l} g_{lk} g_{lm} Q_k^{(n)}(t)\nn\\
        && + \lambda^2 (t) \sum_k L_x \tdiff{g_{km}}{L_x}Q_k^{(n)}(t)\,,
        \nn\\
\eea
where $\lambda(t)=\dot{L}_x(t)/L_x(t)$ and $L_x(t)=c-a(t)$ in our
scenario. The antisymmetric coupling $g_{mk}$ is defined via
\bea
  g_{mk}=L_x\int_{\rm cavity}\pdiff{f_m}{L_x} (\mbox{\boldmath $r$})
        f_k (\mbox{\boldmath $r$}) d^3r
\eea
and is therefore related to the geometry factor $m_{mk}$ via
$g_{mk}=-L_x m_{mk}$. Note that compared to \cite{dalvit} the last
line in (\ref{Edalvit1}) constitutes an additional term, since
in our scenario the coupling between different modes may depend on the
cavity parameters, see also equation (\ref{Egeomfactor}). However,
this difference is of minor relevance, since all these terms
are accompanied by a factor of $\lambda^2(t)$. If one assumes
periodic oscillations of the cavity 
$L_x(t)=L_x[1+\epsilon\sin(\Omega t)+\epsilon f(t)]$, 
these terms can be neglected if the amplitude $\epsilon$ is small. 
(The auxiliary function $f(t)$ is chosen to meet
the continuity conditions on $L_x(t)$, see also \cite{dalvit}.)
Consequently, one can expand
(\ref{Edalvit1}) in powers of $\epsilon \ll 1$ to yield
\bea\label{Edalvit2}
  \ddot{Q}_k^{(n)}(t)+(\Omega_k^0)^2 Q_k^{(n)}(t)
        & = & -2 \Omega_k^0 \pdiff{\Omega_k^0}{L_x} L_x \epsilon
                \sin(\Omega t) Q_k^{(n)}(t)\nn\\
        && -\epsilon\Omega^2\sin(\Omega t)\sum_j g_{kj} 
                Q_j^{(n)}(t)\nn\\
        && +2\epsilon\Omega\cos(\Omega t)\sum_j g_{kj} 
                \dot{Q}_j^{(n)}(t)\nn\\
        && + \epsilon\order{f}+\order{\epsilon^2}\,.
\eea
This equation completely resembles the one found in
\cite{dalvit}. Note however, that we have to use the shifted
eigenfrequencies and the eigenfunctions for leaky cavities. An
approximate solution -- for a more detailed discussion see
\cite{dalvit} -- can be obtained via introducing a new time scale
$\tau=\epsilon t$ and inserting the formal expansion
\bea\label{Edalvit3}
  Q_k^{(n)}(t)=Q_k^{(n)(0)}(t,\tau)+\epsilon Q_k^{(n)(1)}(t,\tau)
        +\order{\epsilon^2}
\eea
with the unknown functions $Q_k^{(n)(0/1)}$ into equation
(\ref{Edalvit2}). Finally, one has to sort in powers of $\epsilon$. 
To lowest order one finds a free harmonic oscillator which can be
solved by
\bea
  Q_k^{(n)(0)}(t,\tau)= A_k^{(n)}(\tau)e^{i\Omega_k^0 t}
                        +B_k^{(n)}(\tau)e^{-i\Omega_k^0 t}\,.
\eea
The next order terms (proportional to $\epsilon$) yield a driven
harmonic oscillator equation for $Q_k^{(n)(1)}$ with the eigenfrequency 
$\Omega_k^0$ 
\bea
  \partial_{t\tau}^2 Q_k^{(n)(1)} + (\Omega_k^0)^2 Q_k^{(n)(1)}
        & = & -2 \partial_{t\tau}^2 Q_k^{(n)(0)}\nn\\
        && - 2 \Omega_k^0 L_x \pdiff{\Omega_k^0}{L_x}\sin(\Omega t)
        Q_k^{(n)(0)}\nn\\
        && - \Omega^2 \sin(\Omega t) \sum_j g_{kj} Q_j^{(n)(0)}\nn\\
        && + 2\Omega\cos(\Omega t) \sum_j g_{kj} \partial_t
        Q_j^{(n)(0)}\,.\nn\\
\eea
In order to keep the expansion (\ref{Edalvit3}) convergent, 
above oscillator must not be at resonance. Consequently, all terms
proportional to $\exp(\pm i \Omega_k^0 t)$ -- with $k$ being the
particular mode of interest -- on the right hand side have to cancel. 
By imposing these conditions for the mode $k=L$ and for the coupling
mode $k=R$ and inserting the frequency deviations
\bea
  \Omega & = &2\Omega_L^0+h=2\Omega_L^0+\epsilon\alpha\,,\\
  \Omega_R^0 & = & 3 \Omega_L^0 + H = 3\Omega_L^0 + \epsilon\beta\,,
\eea
one finds four linear and coupled evolution equations for the
coefficients $A_L^{(n)}(\tau)$, $B_L^{(n)}(\tau)$, $A_R^{(n)}(\tau)$,
and $B_R^{(n)}(\tau)$. These equations are -- apart from the different
couplings and the additional deviation $H$ -- virtually identical with those 
presented in \cite{dalvit}. Having applied the modified phase transformations
\bea
  A_L^{(n)}(\tau) & = & e^{+i\alpha\tau/2} a_L^{(n)}(\tau)\,,\\
  B_L^{(n)}(\tau) & = & e^{-i\alpha\tau/2} b_L^{(n)}(\tau)\,,\\
  A_R^{(n)}(\tau) & = & e^{+3i\alpha\tau/2} e^{-i\beta\tau}
        a_R^{(n)}(\tau)\,,\\
  B_R^{(n)}(\tau) & = & e^{-3i\alpha\tau/2} e^{+i\beta\tau}
        b_R^{(n)}(\tau)
\eea
it is straightforward to rewrite these equations in matrix form.
\bea
  \tdiffonly{\tau} \left( \begin{array}{c} 
                a_k^{(n)} \\  
                b_k^{(n)} \\
                a_j^{(n)} \\  
                b_j^{(n)} \end{array} \right)
                = \underline{M} \left( \begin{array}{c} 
                a_k^{(n)} \\  
                b_k^{(n)} \\
                a_j^{(n)} \\  
                b_j^{(n)} \end{array} \right)\,,
\eea
where
\beann
  \underline{M} = \left( \begin{array}{cccc}
  -\mbox{$\displaystyle\frac{i}{2}$}\alpha & \gamma_1 & \gamma_2 & 0 \\
  \gamma_1 & +\mbox{$\displaystyle\frac{i}{2}$}\alpha & 0 & \gamma_2 \\
  -\gamma_3 & 0 & -\mbox{$\displaystyle\frac{i}{2}$}(3\alpha-2\beta) & 0 \\
  0 & -\gamma_3 & 0 & +\mbox{$\displaystyle\frac{i}{2}$}(3\alpha-2\beta)
                \end{array}\right)\,.
\eeann
The quantities $\gamma_1, \gamma_2, \gamma_3$ have been introduced for
convenience
\bea\label{Enewvalues}
  \gamma_1 & = & \frac{L_x}{2} \pdiff{\Omega_L^0}{L_x} 
        \approx -\frac{1}{2} \frac{(\Omega_L^{x0})^2}{\Omega_L^0}\,,\\
  \gamma_2 & = & \frac{(2\Omega_L^0+h)(2\Omega_L^0+H-h/2)}{2\Omega_L^0}g_{LR}
        \nn\\
           & \approx & 2\Omega_L^0 g_{LR}\,,\\
  \gamma_3 & = & \frac{(2\Omega_L^0+h)(2\Omega_L^0+h/2)}{2(3\Omega_L^0+H)}
           g_{LR}
           \approx \frac{2}{3} \Omega_L^0 g_{LR}\nn\\
           &\approx& \frac{\gamma_2}{3}\,.
\eea

\end{appendix}


\end{document}